\documentclass[12pt]{article}
\usepackage[
top = 2.5cm, 
bottom = 2.5cm,  
left = 2.55cm,  
right = 2.55cm]{geometry}

\usepackage{bbm}
\usepackage{overpic}
\usepackage{subfigure}
\usepackage{caption}
\usepackage{mathtools}
\usepackage{latexsym} 
\usepackage{verbatim}  
\usepackage{tikz}
\usetikzlibrary{matrix}
\usepackage{color}
\usepackage{graphicx,amssymb,amsfonts,amsmath,amssymb,amscd,amstext, mathrsfs}
\usepackage{graphicx}
\usepackage{dsfont}
\usepackage{xcolor}
\usepackage{amsmath}
\usepackage{empheq}
\usepackage{cite}

\numberwithin{equation}{section}

\newlength\dlf

\newcommand{\bw}{\begin{widetext}}
\newcommand{\ew}{\end{widetext}}
\newcommand{\bea}{\begin{eqnarray}}
\newcommand{\eea}{\end{eqnarray}}
\newcommand{\be}{\begin{equation}}
\newcommand{\ee}{\end{equation}}
\newcommand{\nn}{\nonumber}

\renewcommand{\bar}[1]{\overline{#1}}
\renewcommand{\tilde}[1]{\widetilde{#1}}
\renewcommand{\hat}[1]{\widehat{#1}}

\newcommand{\<}{\langle}
\renewcommand{\>}{\rangle}
\newcommand{\norm}[1]{\left\lVert#1\right\rVert}

\newcommand{\CO}{\mathcal{O}}

\newcommand{\Ppeff}{P_+^\textrm{eff}}

\DeclareFontShape{OT1}{cmr}{mx}{n}{<->cmr10}{}
\newcommand{\titlefont}{\fontseries{mx}\selectfont}

\def\frac#1#2{{#1\over #2}}

\usepackage{jheppub}

\begin{document}

\begin{titlepage}

\begin{flushright} 
\end{flushright}

\begin{center} 

\vspace{0.35cm}

{\fontsize{19.5pt}{25pt}
{\titlefont 
Large Momentum EFT and Lightcone Quantization
}}

\vspace{1.6cm}  

{{Hongbin Chen$^1$, A. Liam Fitzpatrick$^1$,  Emanuel Katz$^1$,  Yuan Xin$^{2}$}}

\vspace{1cm} 

{{\it
$^1$Department of Physics, Boston University, 
Boston, MA  02215, USA
\\
\vspace{0.1cm}
$^2$Department of Physics, Yale University, New Haven, CT 06520, USA
}}\\
\end{center}
\vspace{1.5cm}

{\noindent 
We develop methods for computing the effective action at infinite momentum for $1+1d$ QFTs at finite volume which do not rely on the theory having a Lagrangian description.
We do this by taking the infinite momentum limit of equal-time quantization and integrating out all except for the chiral modes of the theory. Our main application of this method is to the Ising Field Theory (IFT), with an energy and magnetic deformation, where we compute the effective lightcone Hamiltonian numerically and check it against results from TCSA. Remarkably, in the low-temperature phase, the Lorentz invariant effective Hamiltonian at infinite momentum takes a very compact form and depends on the volume only through the finite volume vacuum expectation value of $\< \sigma\>$, the spin operator.
}

\end{titlepage}

\tableofcontents

\newpage

\newpage
\section{Introduction and Summary} 
 One can ask the following question regarding Lorentz invariant Quantum Field Theory (QFT):   Is it possible to compute all of its non-perturbative data in manner which respects Lorentz invariance? Haag's theorem suggests that in a Hamiltonian approach one must necessarily compute in finite volume for otherwise one cannot describe the states of the interacting theory in terms of those of the free theory.  Indeed, the overlap between the vacuum of the free theory and the vacuum of the interacting theory is exponentially suppressed with the volume, as the volume becomes large.  This state of affairs seems unsatisfactory as, in practice, in the perturbative regime, we are able to compute using Feynman diagrams, without any reference to the volume, in a manner which retains Lorentz invariance at any order in the couplings.  One may view this as suggesting that some kind of Lorentz invariant approach could be attainable.  Lightcone (LC) Quantization offers the possibility of such an approach as in this framework the vacuum remains trivial and the interacting states can be described directly in terms of the free theory ones.  However, there is a price to pay for this simplification. In modern terms, LC is an effective theory where certain degrees of freedom, in particular `zero modes' with vanishing lightcone momentum $p_- \propto E+p_x =0$, have been `integrated out', potentially generating new contributions to the Hamiltonian.  While these contributions are known in some special cases,\footnote{See e.g.\ \cite{Fitzpatrick:2018xlz,Fitzpatrick:2018ttk, Burkardt:1997bd, Burkardt:1992sz} for previous work on lightcone effective Hamiltonians in scalar Lagrangian theories. } computing them in general remains an open problem, especially because they can reflect nonperturbative information about the theory.  For example, this kind of effective Hamiltonian can have discontinuities in the couplings in the presence of phase transitions.  In fact, in this work we find that some terms in the Hamiltonian are controlled by order parameters.

The goal of the present paper is to explore the structure of the LC effective Hamiltonian by starting with the more standard equal-time quantization framework and then taking the large momentum limit.  In this way, we may compute the LC effective Hamiltonian even for theories with a non-Lagrangian description. The aim of such an exploration is to try and determine the rules of the LC effective theory, ideally so that terms in the Hamiltonian could be written directly in the future.  Besides offering an explicitly Lorentz invariant formulation, having the LC Hamiltonian is also advantageous in practice, in the context of Hamiltonian truncation, as one can reach better resolution in computing observables with far fewer basis states.\footnote{More generally,  several recent works have developed methods to improve the accuracy of Hamiltonian truncation through the use of effective Hamiltonians \cite{Feverati:2006ni,Giokas:2011ix,Lencses:2014tba,Elias-Miro:2017tup,Rychkov:2015vap,Cohen:2021erm,Elias-Miro:2020qwz,EliasMiro:2022pua}}.

We adopt the view that QFTs  are defined as ultraviolet (UV) fixed points, described by Conformal Field Theory (CFT), plus (one or more) relevant deformations that trigger an RG flow. Solving the QFT means solving the Hamiltonian for the CFT plus the deformation:
\begin{equation}
H = H_{\rm CFT} + g \int d^{d-1} x ~\CO(x).
\label{eq:GeneralH}
\end{equation}
Formulated this way, the Hamiltonian is determined purely in terms of CFT data, divorced from the need for a  Lagrangian description.  In two spacetime dimensions, an infinite number of solvable non-Lagrangian CFTs are known, explicitly providing a rich space of such Hamiltonians.  However, if $\CO$ is a primary operator and the deformed theory is gapped, then one can see on general grounds that additional effective interactions will be necessary in LC quantization.   To see why, consider a trial wavefunction for the lightest state with lightcone momentum $p_-$, made from the UV stress tensor $T_{--}$ as follows:
\begin{equation}
| T, p\> \equiv \int d x^- e^{i p_- x^-} T_{--}(x^-) | 0 \>,
\end{equation}
where $|0\>$ is the vacuum,\footnote{In LC, the vacuum is not renormalized, so $|0\>$ is the vacuum of both the original UV CFT and its deformed QFT. } which can be taken to have zero energy without loss of generality, and $x^\pm \propto t\pm x$.  Because of the conservation of $T_{\mu\nu}$ and the tracelessness condition $T_{+-}=0$ in a CFT, $T_{--}$ does not depend on $x^+$ in the UV, so $(P_+)_{ \rm CFT} |T,p\> =0$.  However, if $\CO$ is a primary operator, then it also follows on general grounds that its expectation value in the vacuum, or any Virasoro descendant thereof, vanishes.  Therefore, this trial wavefunction has zero energy in the QFT:
\begin{equation}
\<T, p | P_+ | T, p'\> = g \int dx^- \< T, p | \CO(x) | T, p'\> = 0.
\end{equation}
The energy of the trial wavefunction $|T,p\>$ is an upper bound on the true lightest energy eigenvalue with momentum $p_-$, and therefore in particular it is an upper bound on the mass of the lightest particle, which would imply that theory cannot be gapped.  But this is in contradiction with the fact that there do exist many counterexamples consisting of a CFT deformed by a relevant primary operator that create a gap in the infrared (IR).\footnote{For instance, take a free fermion deformed by a mass term, where the gap is simply the bare fermion mass.}

\begin{figure}
\centering
\includegraphics[width=0.4\linewidth]{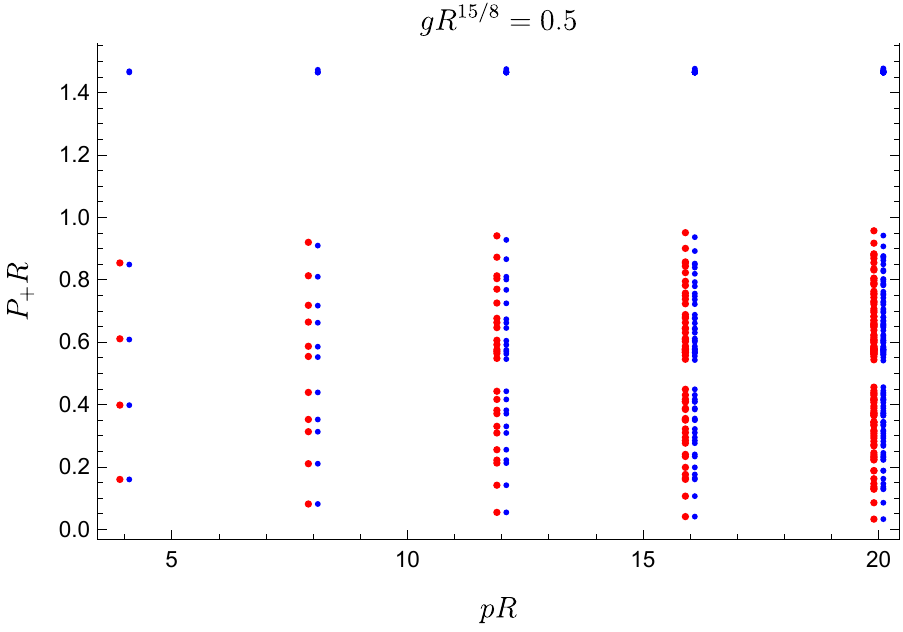}\quad
\includegraphics[width=0.55\linewidth]{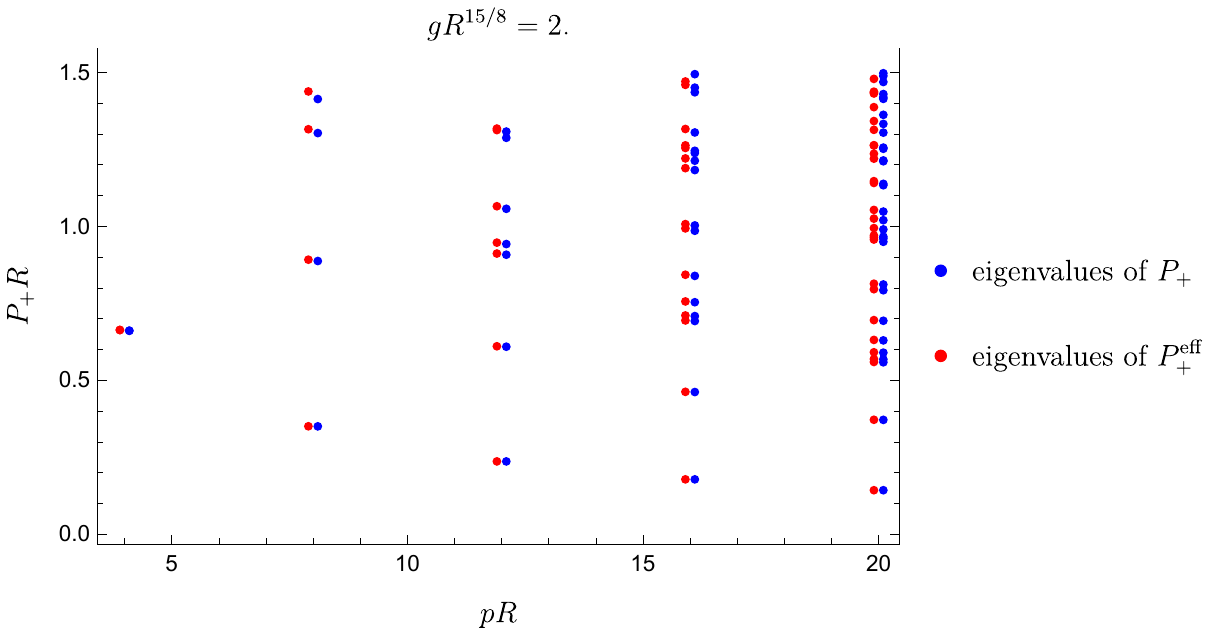}
\caption{Comparison of the spectra of original $P_+$ and $P_+^\textrm{eff}$. Each point in the plots represents one eigenvalue of the corresponding $P_+$. We used $\Lambda R=40$ when making these plots.}
\label{fig:SpectrumComparison}
\end{figure}

Therefore, effective contributions due to integrating out zero modes must be included in order to produce a positive contribution to $\<T, p |P_+^{\rm eff} | T, p'\>$.    In this work 
we develop a numeric method for  computing the LC $P_+^{\rm eff}$ nonperturbatively. We start with the theory in standard equal-time (ET) quantization at finite volume, and take the  Hamiltonian at large spatial momentum $p_x$\footnote{The Hamiltonian truncation formalism for ET quantization is known as the Truncated Conformal Space Approach (TCSA) \cite{Yurov:1989yu,Yurov:1991my,Hogervorst:2014rta,Rychkov:2014eea,James:2017cpc,Horvath:2022zwx}. For a study of TCSA in nonzero momentum frame, see \cite{Chen:2022zms}.}. The large momentum limit of ET is almost the same as LC, the difference being that the latter only keeps a much smaller set of `light' degrees of freedom. In that sense, LC is more economical.
In terms of the ET Hamiltonian $H$, the LC effective Hamiltonian $\Ppeff$ is 
\begin{equation}
\boxed{
\begin{aligned}
\Ppeff= \lim_{p_x \rightarrow \infty} &\frac{1}{\sqrt{2}}\frac{1}{\sqrt{Z}} \left( H_{ll} -p_x- H_{lh} \frac{1}{H_{hh}-p_x} H_{hl}\right)\frac{1}{\sqrt{Z}}, \\
Z& = \mathbbm{1} + H_{lh} \frac{1}{(H_{hh}-p_x)^2} H_{hl},
\end{aligned}
}
\label{eq:HeffIntro}
\end{equation}
where $H$ has been separated into blocks $H_{ll}, H_{hl}$ and  $H_{hh}$ for `light' and `heavy' subspaces as we define in the body of the paper. Knowing the correct `light' degrees of freedom is in general an essential ingredient in formulating any effective theory. We will argue that the correct light degrees of freedom for the large momentum limit of a 2d CFT deformed by a relevant operator are just the chiral modes of the UV CFT.  Directly diagonalizing (\ref{eq:HeffIntro}), which restricts to a basis using only these light chiral modes, and comparing its energy eigenvalues to those of the full theory is a consistency check of this conjecture.  In Fig.~\ref{fig:SpectrumComparison} we present this comparison of eigenvalues between the full and effective theories, for the Ising CFT deformed by the spin operator $\sigma$; in later sections, we will discuss additional requirements that the effective Hamiltonian must satisfy.

In this paper, we apply the formula (\ref{eq:HeffIntro}) to the Ising Field Theory (IFT):
\begin{equation}\label{eq:Hamiltonian}
  H=H_{0}+\frac{1}{2\pi}\int_{0}^{2 \pi R} d x \,\big(  m\varepsilon(x) + g\sigma(x) \big) \, .
\end{equation}
We find that (\ref{eq:HeffIntro}) is able to capture both the low and the high temperature phases of the model.  However, interestingly, in the low temperature phase ($m \leq 0$), numerical evidence suggests that the effective Hamiltonian takes a particularly simple form at large momentum $p$:

\begin{equation}
\Ppeff =  m^2 \int dx^- \psi \frac{1}{2i\partial_-}\psi + \frac{g \langle\sigma\rangle}{2p} M^2_{\sigma} 
+  \CO(\frac{1}{p^2}).
\label{eq:BoostedVolumeDependenceHeffIntro}
\end{equation}
Here, $\langle\sigma\rangle = f(g,m,R)$ is the order parameter, or the finite volume vacuum expectation value of $\sigma$, while $M^2_\sigma$ is an operator that is independent of $g,m,R$, and $p$.\footnote{A qualitatively similar proposal, with $P_+^{\rm eff}$ proportional to the vev of $\cos \beta \phi$ (though strictly at infinite volume), appeared in \cite{Burkardt:1992sz} for the sine-Gordon model. }

Our paper is organized as follows.  In section \ref{sec:Heff}, we motivate and derive the form (\ref{eq:HeffIntro}) for the effective Hamiltonian of the infinite momentum frame and 
describe how standard QFT observables are described in the effective theory.  In section \ref{sec:IsingApp} we apply our general formulation to the Ising Field Theory, first with just the sigma deformation, and then to both the low temperature and high temperature phases.    Then, in section \ref{sec:Dyson} we explain the relation of (\ref{eq:HeffIntro}) to the earlier formulation of the effective Hamiltonian of \cite{Fitzpatrick:2018ttk} in terms of the Dyson series.   Finally, in section  \ref{sec:Discussion}, we conclude by providing an explicit compact expression for $M^2_\sigma$,  and discuss some future directions of investigation.

\section{Effective Hamiltonian}
\label{sec:Heff}
\subsection{Light Degrees of Freedom}

Our goal is to find an effective description of the infinite momentum frame.  A key first step is to identify the appropriate degrees of freedom of the effective theory. Lightcone quantization suggests that we should be able to work with just the chiral degrees of freedom from the UV.  Compared to the original degrees of freedom, this is a significant reduction and it is perhaps surprising that such a small sector of the theory can capture the relevant physics.  However, assuming the full theory is gapped, what is special about the chiral sector is that it has exactly the same structure as the light states at large boost, and so are a natural guess for the effective degrees of freedom.

To see this in more detail, we begin with some standard kinematic facts about the infinite momentum frame, which underly all of our analysis in this paper.  At large momentum, the generator $P_+ \equiv \frac{1}{\sqrt{2}} (H-P)$ develops a light sector.  The reason for this is simply that any particle with mass $\mu$, at $p \rightarrow \infty$  (without loss of generality, we take $p>0$)
\begin{equation}
\begin{aligned}
p_- &= \frac{1}{\sqrt{2}} \left( \sqrt{\mu^2 + p^2} + p \right) \approx \sqrt{2}p + \frac{\mu^2}{2\sqrt{2}p} \\
p_+ &= \frac{1}{\sqrt{2}} \left( \sqrt{\mu^2 + p^2} - p \right) \approx \frac{\mu^2}{2\sqrt{2}p} \, ,
\end{aligned}
\end{equation}
so the $p_+$ eigenvalues of states with large $p >0$ become small.  Moreover, the lightcone momentum $p_-$ is approximately conserved in this limit
\begin{equation}
p_- - p_-' \approx \frac{\mu^2 - \mu'^2}{2\sqrt{2}p}
\end{equation}
just as it is in the lightcone quantization. In the free massless theory, the states with only left-movers form a special sector with $p_+ = 0$. In the infinite boost limit $p \rightarrow \infty$ this sector remains light.  Compared to this light sector, states with right-movers are infinitely heavy and decouple.  Moreover, the vacuum state (with $p_-=0$) decouples due to $p_-$ conservation.  Thus, one sees the kinematics of lightcone quantization emerge naturally from equal-time quantization at infinite momentum.  More concretely, in the infinite momentum limit, a LC effective Hamiltonian operator should satisfy:
\begin{equation}
\label{eq:LCcondition}
p\frac{d}{dp} \langle i,p|\Ppeff|j,p\rangle = -   \langle i,p|\Ppeff|j,p\rangle
\end{equation}
for each matrix element of the Hamiltonian in the proper LC CFT basis of states, $|i,p\rangle$.  This relation insures that the correct Poincare algebra,
\begin{equation}
[J,\Ppeff] = - i \Ppeff,
\end{equation}
is obeyed with $J=\int dx^- x^- T_{--}(x^-)$, the chiral CFT theory boost operator.  Note that it is crucial that the boost operator be the CFT one, as it leaves the CFT vacuum invariant, consistent with LC quantization.  As we shall see, satisfying the requirement (\ref{eq:LCcondition}) can be non-trivial, and in a sense, it characterizes the appropriate low-energy LC degrees of freedom.  In particular, though the low-energy eigenvalues of $P_+$ have the correct scaling with $p$ (as can be seen, for example, in Fig.\ref{fig:SpectrumComparison}), this is not true about all $P_+$ matrix elements in a generic basis.  Obtaining (\ref{eq:LCcondition}) requires first identifying the correct 
low-energy LC degrees of freedom in terms of which $\Ppeff$ can be written.  Our conjecture is that the correct degrees of freedom are the chiral modes, however, even here there can be important subtleties.  For instance, as we shall show, the low temperature phase effective Hamiltonian, (\ref{eq:BoostedVolumeDependenceHeffIntro}), satisfies (\ref{eq:LCcondition}) in the Lightcone Conformal Truncation (LCT) basis  \cite{Katz:2013qua,Chabysheva:2014rra,Katz:2016hxp,Anand:2020gnn}, but not in the Discrete Lightcone Quantization (DLCQ) basis \cite{Pauli:1985ps}.

Our starting point for obtaining the effective lightcone Hamiltonian $\Ppeff$ is to take advantage of the hierarchy of scales that opens up in the large momentum limit.
 A useful strategy for `integrating out' a heavy sector in a Hamiltonian formulation is to decompose the Hamiltonian into the Hamiltonian $H_{hh}$ for the heavy sector and its Schur complement $H/H_{hh}$; this strategy was introduced to Hamiltonian Truncation studies in \cite{Rychkov:2014eea}, and we will follow a similar analysis.  We will use the `Hamiltonian' $P_+ \equiv \frac{1}{\sqrt{2}} (H-P)$ even though we work in ET quantization; the goal is to connect $P_+$ at large momentum to the lightcone effective Hamiltonian.  To begin, we separate $P_+$ into its `CFT' piece $P_+^{(0)}$ and its `deformation' piece $V$:
\begin{equation}
P_+ = P_+^{(0)} + V. 
\end{equation} 
The term $V$ will typically include counterterms.  In particular, it is crucial that any induced vacuum energy is subtracted out, so that the smallest eigenvalue of  $P_+$ vanishes.
At large momentum, the light states 
under $P_+^{(0)}$ are states created by purely chiral operators (with $P_+^{(0)}=0$). 
Define projection operators $\Pi_l$ and $\Pi_h$ onto the space of such states and the space perpendicular to them, respectively:
\begin{equation}
\Pi_l \equiv \sum_{l \in L} | l \> \< l | , \qquad \Pi_h  \equiv \sum_{h \notin L} | h \> \< h | , \qquad \mathbbm{1} = \Pi_l + \Pi_h,
\end{equation}
where $L$ is a set of basis states for the `light' states of $P_+^{(0)}$. When the UV CFT is that of a free fermion,  a basis for $L$ is the usual Fock space basis for states made of only left-movers.

A general state $| \psi\>$ can be decomposed into its light and heavy components:
\begin{equation}
|\psi \> = (\Pi_h + \Pi_l) | \psi\> = |\psi\>_h + |\psi\>_l , \qquad | \psi\>_l = \Pi_l | \psi\> , \quad |\psi\>_h = \Pi_h | \psi\>.
\end{equation}
Next, we define an orthonormal basis $| \hat{n} \>$  of energy eigenstates of the interacting Hamiltonian $P_+$
\begin{equation}
P_+ | \hat{n} \> = E_n | \hat{n} \>.
\end{equation}
We can break up each eigenstate $|\hat{n}\>$ into its $h$ and $l$ components to write
\begin{equation}
P_+ | \hat{n}\>_h + P_+ |\hat{n}\>_l = E_n | \hat{n}\>_h + E_n | \hat{n}\>_l 
\label{eq:HamEqComponents}
\end{equation}
Restricted to the `heavy' components, this equation says
\begin{equation}
(P_+)_{hh} |\hat{n}\>_h + (P_+)_{hl} | \hat{n}\>_l = E_n | \hat{n}\>_h\Rightarrow  |\hat{n}\>_h = -\frac{1}{(P_+)_{hh} -E_n} (P_+)_{hl} |\hat{n}\>_l ,
\label{eq:HeavyFromLight}
\end{equation}
where we have broken up $P_+$ into four `blocks', 
\begin{equation}
P_+ = \left( \begin{array}{cc} (P_+)_{ll} & (P_+)_{lh} \\ (P_+)_{hl} & (P_+)_{hh} \end{array} \right).
\end{equation}
Now, note that equation (\ref{eq:HeavyFromLight}) is a formula for the `heavy' components of an eigenstate in terms of its light eigenstates.  For a generic eigenstate, this equation is not particularly useful for our purposes because it also requires knowing the eigenvalue $E_n$.  However, in the special case where $E_n$ is small, we can expand
\begin{equation}
 |\hat{n}\>_h = -\frac{1}{(P_+)_{hh}} (P_+)_{hl} | \hat{n}\>_l + \CO(E_n) .
 \label{eq:HeavyFromLightComponents}
 \end{equation}
More precisely,  the subleading $\CO(E_n)$ terms are suppressed by powers of $E_n (P_+)_{hh}^{-1}$, so we need $E_n$ to be small compared to the smallest eigenvalue of $(P_+)_{hh}$ in order to perform this expansion.  In the large momentum limit, there will be an infinite number of eigenvalues approaching zero like $E_n \sim 1/|p|$, for which the subleading terms vanish.  

To derive the effective Hamiltonian, next consider the `light' components of (\ref{eq:HamEqComponents}):
\begin{equation}
(P_+)_{lh} | \hat{n}\>_h + (P_+)_{ll} |\hat{n}\>_l = E_n | \hat{n}\>_l \Rightarrow \left( (P_+)_{ll} - (P_+)_{lh} \frac{1}{(P_+)_{hh}- E_n} (P_+)_{hl} \right) | \hat{n}\>_l = E_n |\hat{n}\>_l.
\end{equation}
For small eigenvalues $E_n \ll (P_+)_{hh}$,\footnote{More precisely, for $E_n$ small compared to the smallest eigenvalues of $(P_+)_{hh}$, as mentioned earlier.  } it is tempting to discard the factor of $E_n$ in the denominator.  However, this is not allowed.  The reason is that, as can be seen manifestly from the RHS of the equation, the {\it leading} order part of the expression is $\CO(E_n)$, and only $\CO(E_n^2)$ and higher terms are subleading.  Instead, when we expand in powers of $E_n$, we must keep the $\CO(E_n)$ terms, to get:
\begin{equation}
\left( (P_+)_{ll} - (P_+)_{lh} \frac{1}{(P_+)_{hh}} (P_+)_{hl} \right) | \hat{n}\>_l = E_n (\mathbbm{1}+ \Delta Z) |\hat{n}\>_l + \CO(E_n^2),
\label{eq:GeneralizedEvalEqn}
\end{equation}
where we define
\begin{equation}
 \Delta Z \equiv  (P_+)_{lh} \frac{1}{(P_+)^2_{hh}} (P_+)_{hl}, \qquad Z \equiv \mathbbm{1} + \Delta Z.
 \label{eq:Zdef}
 \end{equation}
Note that $Z$ is defined so that it acts only within the space of `light' components, and not on the full Hilbert space. Discarding the $\CO(E_n^2)$ terms, we obtain a generalized eigenvalue equation for the light components $|\hat{n}\>_l$ of the light eigenstates $|n\>$ of $P_+$.  A drawback of this equation is that the eigenvectors $| \hat{n}\>_l$ are {\it not} orthonormal.  The reason is that we started with the full orthonormal basis states $|\hat{n}\>$ of $P_+$ and projected them onto their light components $|\hat{n}\>_l$, but they will not typically be orthonormal after this projection.  However, using the relation (\ref{eq:HeavyFromLightComponents}) between the heavy and light components, we can easily work out the Gram matrix of the projected eigenstates $|\hat{n}\>_l$:
 \begin{equation}
 \delta_{nm} = \< \hat{m} | \hat{n}\> 
 = \left( {}_l \< \hat{m} | + {}_h \< \hat{m} | \right) \left( | \hat{n} \>_h + | \hat{n} \>_l\right) 
  = {}_l \< \hat{m} | Z | \hat{n} \>_l  .
 \end{equation}
 Therefore, we can simply use the matrix $Z$ to construct a new set of orthonormal basis states $|\tilde{n}\>$ out of the light components of the eigenstates:
 \begin{equation}
 |\tilde{n} \> \equiv Z^{\frac{1}{2}} | \hat{n}\>_l \Rightarrow \< \tilde{m} | \tilde{n} \> = \delta_{nm}.
 \ee
We define $Z^{\frac{1}{2}}$ as the unique symmetric positive definite square root matrix of $Z$ (which is itself symmetric and positive definite). Finally, writing our generalized eigenvalue equation (\ref{eq:GeneralizedEvalEqn}) in terms of the orthonormal states $|\tilde{n}\>$, we obtain (discarding $\CO(E_n^2)$ terms)
\begin{equation}
\Ppeff | \tilde{n} \> = E_n | \tilde{n}\> ,
\end{equation}
where our effective Hamiltonian is
\begin{equation}
\boxed{\Ppeff \equiv Z^{-\frac{1}{2}} \left( (P_+)_{ll} - (P_+)_{lh} \frac{1}{(P_+)_{hh}} (P_+)_{hl} \right)  Z^{-\frac{1}{2}} . }
\label{eq:EffectiveHamiltonian}
\end{equation}
Leading large momentum $p$ corrections to this formula can be included to improve the convergence rate in $1/p$. We relegate the details of how to compute these corrections to appendix \ref{eq:PeffCorrections}. The effective Hamiltonian with the first $1/p$ correction included is
\begin{equation}
\Ppeff+ \delta \Ppeff, 
\label{eq:deltaPpeff}
\end{equation}
with $\delta \Ppeff$ given in equation (\ref{eq:EffectiveHamiltonianCorrection}).

It will be useful to relate the matrix $Z$ to the overlap $\< l |\hat{n}\>$ between eigenstates $|l \>$ of $P_+^{(0)}$ and eigenstates $|\hat{n}\>$ of $P_+$. To this end, start with the relation $|\hat{n}\>_l= Z^{-\frac{1}{2}} |\tilde{n}\>$ and insert a complete set of `light' states $|l\>$:
\begin{equation}
W_{nl} \equiv \< \hat{n} | l \> = {}_l \<\hat{n} | l \> = \< \tilde{n} | Z^{-\frac{1}{2}} | l \> = \sum_{l'} \<\tilde{n} | l' \> \< l' | Z^{-\frac{1}{2}} | l \> = \sum_{l'} (S_{\rm eff})_{n l'} \< l' | Z^{-\frac{1}{2}} | l \>.
\label{eq:Wdef}
\end{equation}
We can write this relation schematically as
\begin{equation}
W = S_{\rm eff} Z^{-\frac{1}{2}}.
\label{eq:WSZ}
\end{equation}
We have introduced the matrix $(S_{\rm eff})_{nl'} \equiv \< \tilde{n} | l'\>$, which is the change of basis between the `light' basis states $|l\>$ and the eigenstates of $\Ppeff$.  In other words, if we write $\Ppeff$ in the $|l\>$ basis, then $S_{\rm eff}$ is the matrix that diagonalizes it:
\begin{equation}
\< l | \Ppeff | l'\> = \sum_{E_n \in \{ \Ppeff \textrm{ eigenvalues}\}} (S_{\rm eff}^\dagger)_{l n} E_n (S_{\rm eff})_{n l'}.
\label{eq:PpEffProj}
\end{equation}

Because of momentum conservation, $P_+$ is block diagonal with different momentum sectors, and (\ref{eq:EffectiveHamiltonian}) should be understood to be the effective Hamiltonian for an individual momentum sector.  In order for the eigenvalues $E_n$ to be small, the momentum $p_x$ must generally be large, with one important exception: the rest frame $p_x=0$.  In the rest frame, the smallest eigenvalue vanishes by construction, because the deformation $V$ is defined to include a counterterm that sets the energy of the interacting vacuum state $|\Omega\>$ to zero.  In this case, the only `light' basis state is just the free theory Fock space vacuum $|0\>$, with $P_+^{(0)} | 0\> =0$.  The `matrix' $\Ppeff = Z^{-\frac{1}{2}} ((P_+)_{ll} - (P_+)_{lh} (P_+)_{hh}^{-1} (P_+)_{hl}) Z^{-\frac{1}{2}} $ is just a number, which is exactly zero by construction, and the $Z$ factor is given by the overlap between the true vacuum and Fock space vacuum:
\begin{equation}
(Z_{p_x=0})^{-1} = | \< 0 | \Omega\>|^2.
\label{eq:Zvac}
\end{equation}
We have explicitly written the dependence of $Z$ on the momentum sector for emphasis.

\subsection{Effective Operators and the Stress Tensor}

Using the formalism in the previous subsection, we can go farther and also calculate how local operators from the UV theory act on the space of states in the low energy description.  The key point is that for light states $|\tilde{n}\>= Z^{\frac{1}{2}} |\hat{n}\>_l$, their corresponding vector in the full theory is fixed by (\ref{eq:HeavyFromLightComponents}).  

Consider an operator $\CO$ defined in the original UV CFT.  Just as we did with the Hamiltonian itself, we can separate $\CO$ into different blocks for the heavy and light sectors:
\begin{equation}
\CO = \left( \begin{array}{cc} \CO_{ll} & \CO_{lh} \\ \CO_{hl} & \CO_{hh} \end{array} \right).
\end{equation}
We want to define an ``effective'' version $\CO^{\rm eff}$ of this operator with the property that $\CO^{\rm eff}$ acting on the light states $|\tilde{n}\>$ is the same as the original UV operator $\CO$ acting on the corresponding eigenstates of the original Hamiltonian:
\begin{equation}
\< \tilde{n} | \CO^{\rm eff} | \tilde{n}'\> \equiv \< \hat{n} | \CO | \hat{n}'\>.
\label{eq:OeffDefn}
\end{equation}
Separating $|\hat{n}\rangle$ into the light sector and the heavy sector
\begin{equation}
	|\hat{n}\rangle = |\hat{n}\rangle_l +|\hat{n}\rangle_h,
\end{equation} 
and using (\ref{eq:HeavyFromLightComponents}) together with $|\tilde{n} \> = Z^{\frac{1}{2}} |\hat{n}\>_l$, we find
that the correct expression for $\CO^{\rm eff}$ is 
\begin{equation}
\CO^{\rm eff} = Z_{p_{\rm in}}^{-\frac{1}{2}} \left( \CO^{ll}-P_+^{lh} \frac{1}{P_+^{hh}} \CO^{hl}- \CO^{lh} \frac{1}{P_+^{hh}} P_+^{hl}+P_+^{lh} \frac{1}{P_+^{hh}}\CO^{hh}\frac{1}{P_+^{hh}}P_+^{hl}\right) Z_{p_{\rm out}}^{-\frac{1}{2}}.
\end{equation}
A particularly important case will be played by the stress tensor, and especially its $T_{--}$ components.  Because $T_{--} \sim \psi \partial_- \psi$ is chiral, it does not have any $T_{--}^{hl}$ or $T_{--}^{lh}$ components, so we can write $T_{--}^{\rm eff}$ more compactly as
\begin{equation}
	\left\langle l\left|T_{--}^{\mathrm{eff}}\right| l^{\prime}\right\rangle=\left\langle l\left|Z_{p_{\rm in}}^{-\frac{1}{2}}(T_{--}^{ll}+\frac{P_+^{lh}}{P_+^{hh}}T_{--}^{hh}\frac{P_+^{hl}}{P_+^{hh}})Z_{p_{\rm out}}^{-\frac{1}{2}}\right|l^{\prime}\right\rangle .
	\label{eq:TmmEff}
	\end{equation}

In standard lightcone quantization, $T_{--}$ is not affected by the interaction, so one might expect that at infinite boost only the ``free'' piece $T_{--}^{ll}$ would contribute to $T_{--}^{\rm eff}$. Moreover, acting on chiral states, $\int dx^- T_{--}$ is already equal to the total lightcone momentum $P_-$ in the UV CFT, again suggesting it does not get renormalized in the LC effective theory. That is, 
\begin{equation}
\int dx^- T_{--}^{\rm eff} = \int dx^- T_{--}^{ll},
\label{eq:TmmIntegratedCondition}
\end{equation}
since both the LHS and the RHS are the momentum $P_-$.  So, it is natural to conjecture that 
\begin{equation}
T_{--}^{\rm eff} = T_{--}^{ll}
\label{eq:TmmNonRenorm}
\end{equation}
 in the large momentum limit. In Fig.~\ref{fig:NoTmmRenorm} we show an explicit numeric check of this nonrenormalization at large $p$ for the case of the Ising model deformed by the magnetic deformation $\sigma$, with coupling $g R=1$ (see (\ref{eq:Hamiltonian}) for our conventions).\footnote{Moreover, we will see  that at large $p$, the effective Hamiltonian in the case of a single relevant deformation depends on $R$ only through an overall  prefactor, which implies that its eigenvectors are independent of $R$.  But  $T_{--}$ is one of our basis states, so its overlap with the physical eigenstates is just one of the coefficients of their eigenvectors.  Therefore in this case $T_{--}$ itself is independent of $R$. }  
 
\begin{figure}
 \center
 	 \includegraphics[width=0.7\textwidth]{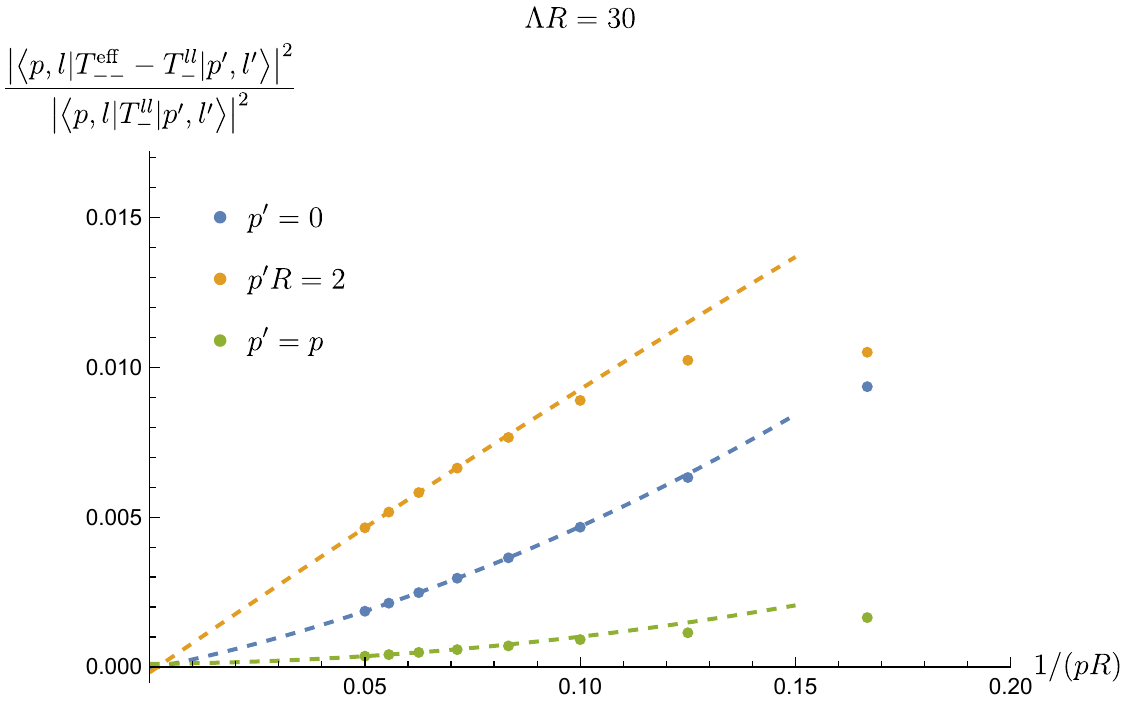}
 	 \caption{Difference of $T_{--}^{\mathrm{eff}}$ and $T_{--}^{l l}$ as a function of $1/p$, for Ising Field Theory  at $gR=1, m=0$ (see \ref{eq:Hamiltonian})). The norm of the matrices are defined as $|M|^2=\sum_{i,j=1}^n M_{ij}^2$. }  
	 \label{fig:NoTmmRenorm}
 \end{figure}

 To sharpen this argument, note that (\ref{eq:TmmIntegratedCondition}) implies
 \begin{equation}
 T_{--}^{\rm eff} = T_{--}^{ll} + \partial_- K
 \end{equation}
 for some  operator $K$ with $\int dx^- \partial_- K =0$.  But by definition, both $T_{--}^{\rm eff}$ and $T_{--}^{ll}$ are operators that act only within the chiral basis, so we can act with the LHS and RHS of the equation above on the vacuum to get an equality between two low-energy states, $|T^{\rm eff}_{--} \> = |T^{ll}_{--}\> + | \partial_- K\>$.  Moreover, $|\partial_- K\>$ has no overlap with $|T_{--}^{ll}\>$, since any such overlap would contribute to $\< T_{--}^{ll} \int dx^- \partial_- K\>$ which vanishes by definition.    Finally, both $|T^{\rm eff}_{--} \> $ and $|T^{ll}_{--} \> $ have the same normalization, since they both measure the momentum of states.  Altogether, these observations imply that
 \begin{equation}
0 =  \norm{|T_{--}^{\rm eff}\>}^2- \norm{|T_{--}^{ll}\>}^2 = \norm{|\partial_- K \>}^2,
 \end{equation}
 so that $|\partial_- K \>$ vanishes.

This `nonrenormalization' of $T_{--}$ will be extremely useful, because it will allow us to perform matching computations between the boosted TCSA  and the effective theory by computing expectation values and correlators of $T_{--}$ in both descriptions and comparing them.

\section{Application to Ising Field Theory}
 \label{sec:IsingApp}
In this section we apply the above analysis to the Ising Field Theory. In previous work \cite{Chen:2022zms} we studied IFT at finite boost and found that working in a boosted frame improved the accuracy of the numeric results. Here, we begin with the same boosted Hamiltonian, and then follow the procedure outlined in the previous sections in order to integrate out the non-chiral modes and obtain a lightcone effective Hamiltonian (\ref{eq:EffectiveHamiltonian}). In principle, we would like to work at infinite $p$, however, in this paper we will simply evaluate (\ref{eq:EffectiveHamiltonian}) by brute force, so $p$ will be large but finite.  At finite $p$ the leading order effective Hamiltonian (\ref{eq:EffectiveHamiltonian}) has finite $p$ corrections, and in order to improve our accuracy in Appendix \ref{sec:effectiveHamiltonainCorrection} we derive the next order correction in $1/p$ (\ref{eq:EffectiveHamiltonianCorrection}).

The Hamiltonian of  IFT is 
\begin{equation}\label{eq:Hamiltonian}
  H=H_{0}+\frac{1}{2\pi}\int_{0}^{2 \pi R} d x \,\big(  m\varepsilon(x) + g\sigma(x) \big) \, .
\end{equation}
with
\begin{equation}\label{eq:IsingCFTH}
  H_{0}=\frac{1}{2 \pi} \int_0^{2\pi R}dx(\psi \bar{\partial} \psi+\bar{\psi} \partial \bar{\psi}).
\end{equation}
The infinite volume limit is reached via taking $R\rightarrow \infty$, or equivalently $m, g \rightarrow \infty$ in units where $R=1$.
In the continuum limit, the theory depends only on the dimensionless parameter 
\begin{equation}
\eta = \frac{m}{h^{8/15}},\qquad \textrm{with } g=2\pi h.
\end{equation}
And we define a dimensionful quantity
\begin{equation}\label{dimensionlessVolume}
t = \sqrt{ m^2 + h^{\frac{16}{15}}}
\end{equation}
such that the dimensionless combination $t R$ indicates the volume.
The IFT Hamiltonian (\ref{eq:Hamiltonian}) describes an Ising CFT (a massless free fermion) deformed by an energy density operator (i.e. a mass term) $\varepsilon$ and a magnetic field operator (i.e. a twist field) $\sigma$. In the limit $m=0$, the theory is the integrable $E_8$ affine Toda theory \cite{Zamolodchikov:1989fp}, and the limit $g=0$ is the massive free fermion theory. The intermediate $\eta$ cases are non-integrable but can be studied numerically via Hamiltonian truncation.  In fact, the most accurate results available come from Hamiltonian truncation in a boosted frame, so the most stringest test we can perform is to compare the results of our effective Hamiltonian to those of the original boosted Hamiltonian, and in addition to make this comparison at different values of the momentum. 

\subsection{Spin Deformation Only (\texorpdfstring{$\eta=0$}{eta=0}) }
\label{sec:IsingSigma}

\subsubsection{Lorentz-invariance and Volume-Independence at Large \texorpdfstring{$p$}{p}}
In a typical Hamiltonian framework for QFT of the form (\ref{eq:GeneralH}), the large volume limit is a strongly coupled limit.  
The reason is that observables depend on dimensionless quantities, and in particular in the rest frame they depend on volume $2 \pi R$ times the coupling $g$ to a power $\nu = \frac{1}{2-\Delta}$ fixed by dimensional analysis. Since  $g^\nu R$ is the only dimensionless free parameter, one is therefore forced to choose between perturbation theory, where it is small, and the large volume limit, where it is large. More precisely, there are dynamical `particle going around the world' finite volume corrections that are parametrically suppressed by $\sim e^{-2 \pi R m_{\rm gap}}$, which only become small when $m_{\rm gap} R \sim g^\nu R$ is large.\footnote{There are also kinematic corrections that are suppressed by power laws in $1/R$, due to the quantization condition on momenta in finite volume.  } 
Remarkably, we will find that  these finite volume corrections simplify dramatically in the large momentum limit, and consequently make it possible to study the large volume limit even at small $g$.\footnote{A similar relation between the small- and large-volume limits appears to be important for the study of the BFSS model \cite{Banks:1996vh}, see \cite{Guijosa:1998rq} for a discussion and references. For another perspective, in a large volume expansion, many finite volume effects can be seen explicitly to vanish at large momentum using the methods from \cite{Luscher:1985dn}.}  In this section, we will restrict our attention to the special case with $m=0$ (i.e., $\eta=0$).  In the limit of large momentum $p$, after fixing the energy counterterm in the ET Hamiltonian, the effective Hamiltonian (\ref{eq:HeffIntro}) for the light states takes the form (\ref{eq:BoostedVolumeDependenceHeffIntro}) from the introduction:
\begin{equation}
\Ppeff =  \frac{g f(g,0,R)}{2p } M^2_\sigma 
+  \CO(\frac{1}{p^2}),
\label{eq:BoostedVolumeDependenceHeff}
\end{equation}
with a volume-dependent factor $f$, which typically must be obtained numerically, and a volume-and momentum-independent operator $M^2_\sigma$. Moreover, we will empirically find that $f$ is simply the volume-dependent vev of $\sigma$.

\subsubsection{Numerical Results}

In order to verify the form of (\ref{eq:BoostedVolumeDependenceHeff}) we must determine the energy counterterm at large momentum.  We will use the Ward Identity to fix the boosted frame vacuum energy (see Appendix \ref{app:Vac}).\footnote{In practice, we have found that using the Ward identity is more accurate for extracting the vacuum energy than computing it in the rest frame with TCSA.  Because the $P_+$ energy of boosted states is small, $\sim 1/p$, small errors in the vacuum energy can lead to significant errors in the lightcone energy. }   This is shown in Fig.~\ref{fig:EvacSigOnly}, where we also compare to exact results from integrability \cite{Klassen:1990dx}.  Note that $E_{\rm vac}/g^2$ vanishes at large volumes, but approaches approximately $- 8$ at small volumes.  Both these limits can be easily understood analytically. By dimensional analysis, at large coupling $g$ the vacuum energy scales as $E_{\rm vac} \sim g^{\frac{16}{15}}$ (in fact, from integrability results it is $E_{\rm vac}= -0.169 g^{\frac{16}{15}}$); therefore $E_{\rm vac}/g^2$ vanishes at large $g$.  At small $g$, the $\CO(g^2)$ contribution from the Ramond vacuum $|\sigma\>$ is just $g^2$ divided by the energy denominator $E_R - E_{NS}$ in the CFT limit, which is just $\Delta_\sigma = 1/8$ (in units with $R=1$).  The contribution from excited Ramond sector states is small, and can be computed either using conformal perturbation theory (\ref{perturbativeEvac}) or integrability methods; these corrections increase the coefficient of $g^2$ from $-8$ to $-\frac{2 \Gamma \left(\frac{1}{16}\right)^2 \Gamma \left(\frac{15}{8}\right)}{7 \Gamma
   \left(\frac{1}{8}\right) \Gamma \left(\frac{15}{16}\right)^2} = -8.00949$, whereas the numerical value we obtain from imposing the Ward identity is $-8.00934$.

With the vacuum energy removed, the only remaining ingredient is that of the coefficient of the $M^2_\sigma$ operator itself, the function $f(g,0,R)$ in (\ref{eq:BoostedVolumeDependenceHeff}). We can numerically determine this function by computing the mass gap as a function of $g$; this is shown in the left panel of Fig.~\ref{fig:SpectrumSigOnly}.  Moreover, 
the form of (\ref{eq:BoostedVolumeDependenceHeff}) dictates that at large momentum, dimensionless quantities should become volume independent. This is indeed what we see for the mass ratios in the second and third panels of Fig.~\ref{fig:SpectrumSigOnly}.

  \begin{figure}
 \centering
 	 \includegraphics[width=0.4\linewidth]{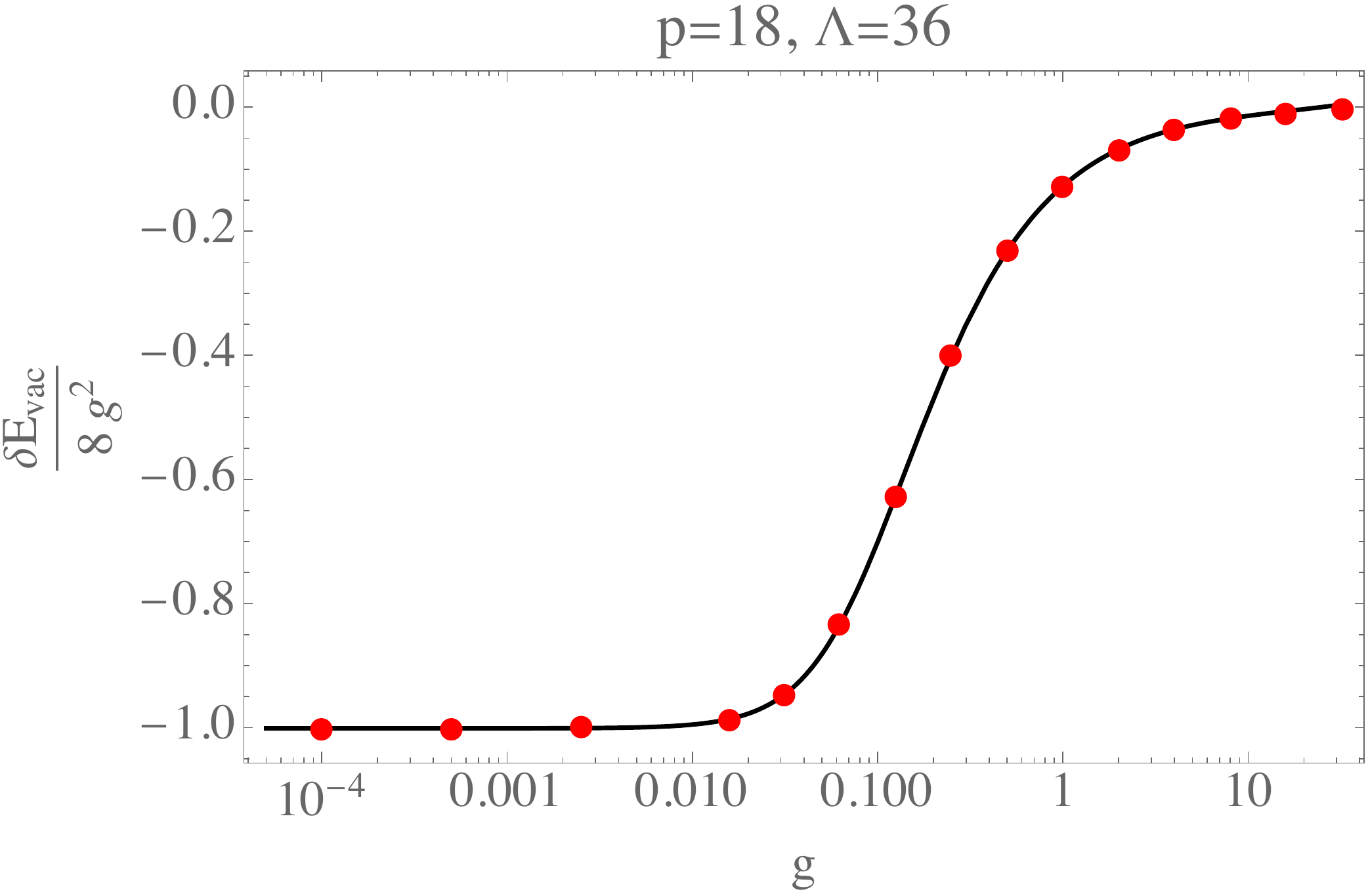}	 
 	 \caption{Vacuum energy shift $\delta E_{\rm vac} \equiv E_{\rm vac} + \frac{c}{12}$ from interactions as a function of coupling $g$, in units of $R=1$.  {\it (Dots, Red)}: Vacuum energy fixed by Ward identity.  {\it (solid, black)}: Vacuum energy from integrability results \cite{Klassen:1990dx}. }
	 \label{fig:EvacSigOnly}
 	  \end{figure}

\begin{figure}
\centering
\includegraphics[width=0.3\textwidth]{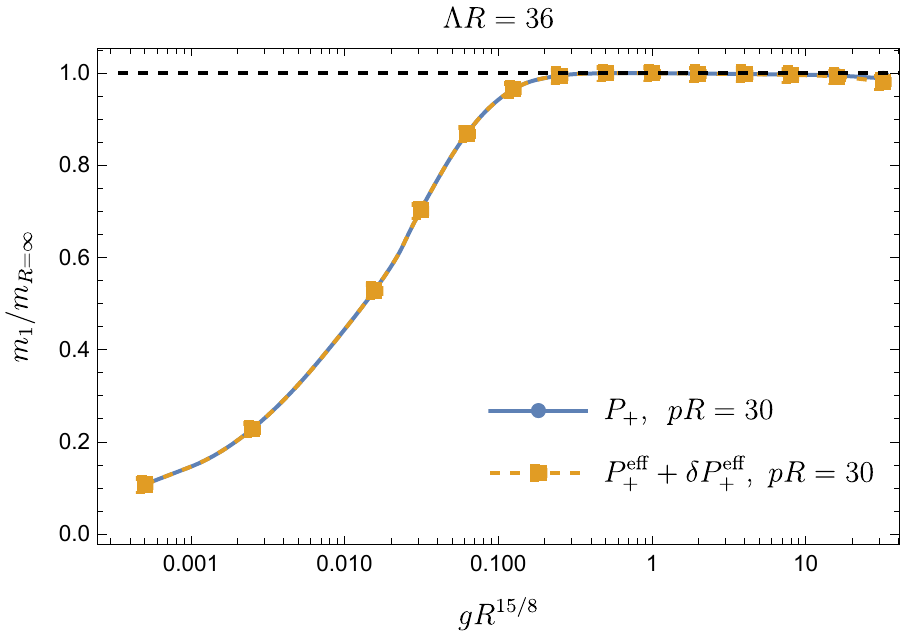}\quad
\includegraphics[width=0.3\textwidth]{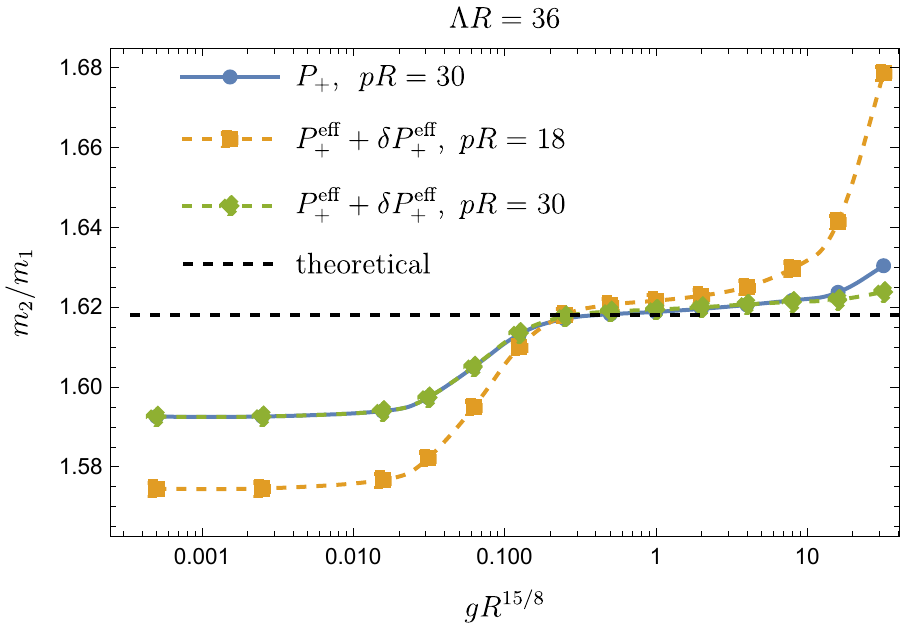}\quad
\includegraphics[width=0.3\textwidth]{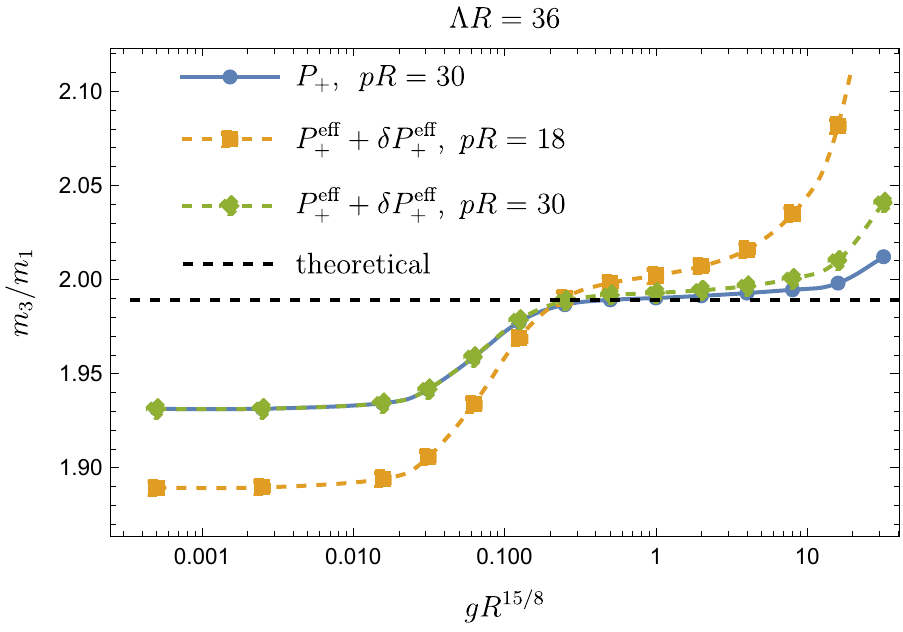}
\caption{Left: mass gap computed numerically compared with the infinite volume mass gap $m_{R=\infty}\simeq 4.40490858 (g/(2\pi))^{8/15}.$ Middle and right: the ratios of the second and third bound states with the first bound states compared with the theoretical values: $m_2/m_1\simeq 1.61803, m_3/m_1 = 1.98904$.}
\label{fig:SpectrumSigOnly}
\end{figure}

The $c$-function contains additional dimensionless observables.  In particular, there are three isolated stable states below the multiparticle continuum, and we will focus on their contributions $c_1, c_2, c_3$ to the $c$-function.  These are shown in Fig.~\ref{fig:cfuncSigOnly}, where we compute both the result from the original boosted Hamiltonian $P_+$ as well as the effective Hamiltonian, including the leading $1/p$ correction $\delta P_+^{\rm eff}$.  The exact theoretical predictions for the $c_i$s from integrability are also shown for comparison.  The first point to note is that the agreement is extremely good over a wide range of couplings, up until about $g R^{15/8} \sim 10$ for $c_1$ and up to $g R^{15/8} \sim 1$ for $c_2$ and $c_3$.  Moreover, the agreement improves with increasing $p$.  In all cases, we have taken the truncation cutoff to be $\Lambda R=36$.  Due to the orthogonality catastrophe, going to larger values of $g$ requires increasing the cutoff $\Lambda$.  However, a remarkable feature of the numeric results is that the predictions for the $c_i$s agree with the theoretical values over a range of volumes, even as the volume becomes very small.  This is yet another indication that at large momentum, one gets the structure of (\ref{eq:BoostedVolumeDependenceHeff}).

\begin{figure}
\centering 
\includegraphics[width=0.3\textwidth]{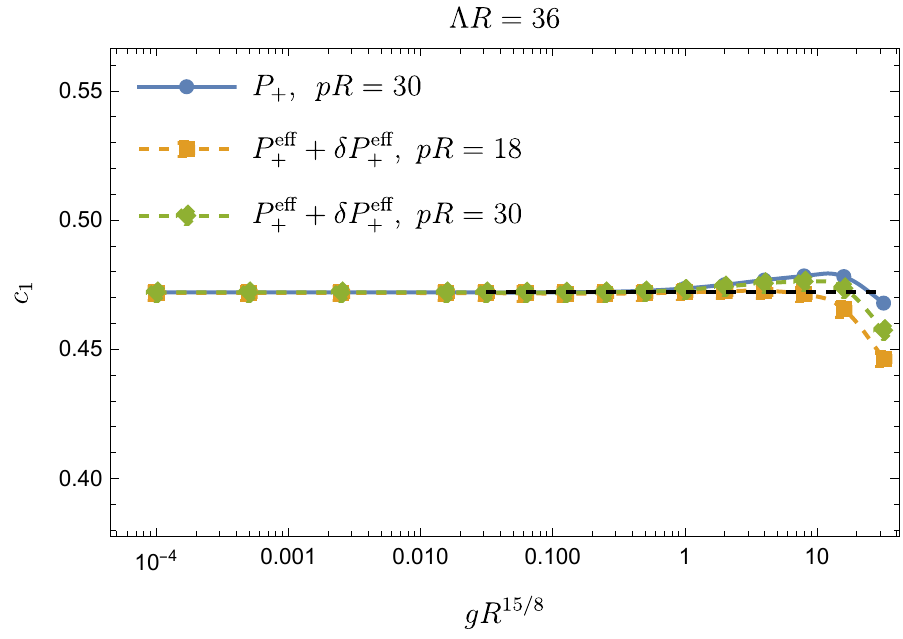}
\includegraphics[width=0.3\textwidth]{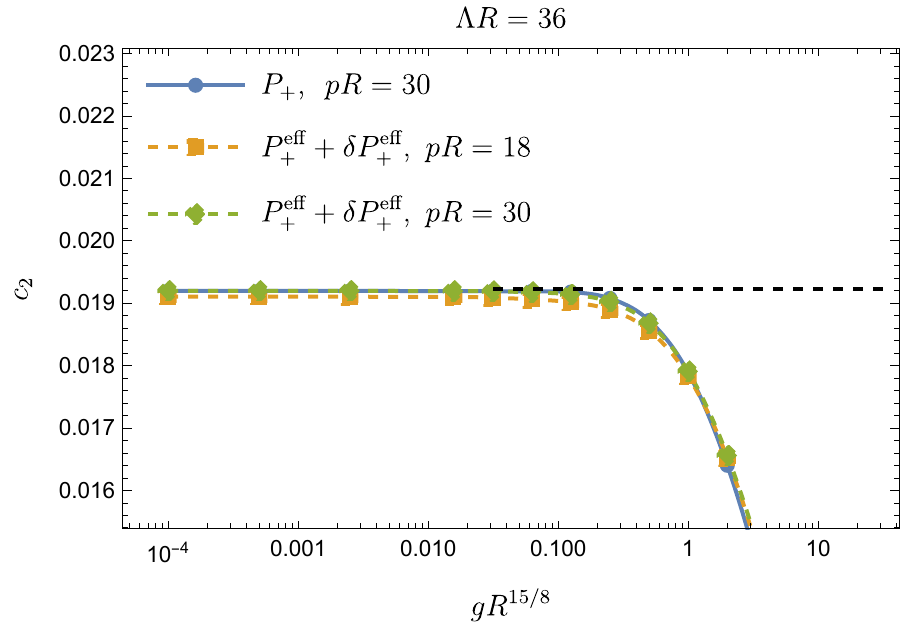}
\includegraphics[width=0.3\textwidth]{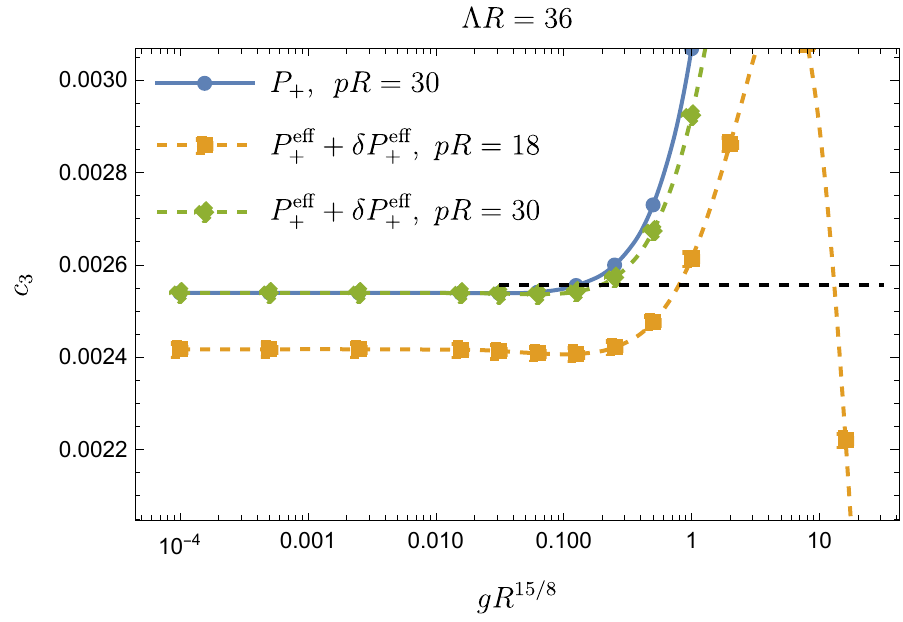}
\caption{ $c_1$ computed from $T_{--}$. For $P_+$, we used the interacting vacuum $|\Omega\rangle$, while for the $P_+^\textrm{eff}$, we used the Fock-space vacuum and $T_{--}^{ll}$.  The vacuum energy is fixed by using the Ward identity.}
\label{fig:cfuncSigOnly}
	\end{figure}

Thus far, we have checked (\ref{eq:BoostedVolumeDependenceHeff}) indirectly by computing various dimensionless observables.  However, it is also interesting to ask whether one can compute the effective Hamiltonian directly for different volumes at large momentum.  That is, we would like to compute the operator $M^2_\sigma$ in some basis and verify that it is volume independent.  Note that the form of (\ref{eq:BoostedVolumeDependenceHeff}) implies that in the appropriate basis, the effective Hamiltonian must obey the condition (\ref{eq:LCcondition}) for all its matrix elements.  In other words, all matrix elements must go to zero at large momentum.  Though we expect that such a LC Hamiltonian must exist for chiral states, there can be important subtleties.  In particular, not all chiral wavefunctions have finite matrix elements for a given LC Hamiltonian.  A simple example of this 
is the case of a free fermion, where the effective fermion mass term $\psi \frac{1}{i\partial_-} \psi$ diverges when the momentum fraction $x$ of any individual parton vanishes, due to the $1/\partial_-$ factor.  In \cite{Katz:2016hxp}, the space of wavefunctions that vanish at $x=0$, and therefore have finite energy in the presence of this mass term, was called the `Dirichlet subpace'.  From our construction of the original Hamiltonian, we can check explicitly which sets of chiral states remain light.  Interestingly, in the presence of the $\sigma$ deformation, we find that the standard DLCQ states are lifted out of the spectrum, as shown in Figure \ref{fig:HeffME}.  What is special about the DLCQ states is that they have wavefunctions that are essentially delta-function localized at a definite value of parton-$x$ for each parton.  By contrast, chiral states whose wavefunctions in the free fermion basis are smooth wavefunctions over the individual parton fractions appear to remain in the low energy spectrum.  Consequently, to verify (\ref{eq:BoostedVolumeDependenceHeff}) directly we cannot use DLCQ or the discrete Fock space states at large quantized momentum.  Instead, we will compute $M^2_\sigma$ matrix elements in the Conformal Truncation (LCT) basis, which are a set of smooth polynomials. In this basis (\ref{eq:LCcondition}) is satisfied for the $\sigma$ deformation.  

\begin{figure}[htbp] 
\begin{center}
\includegraphics[width=0.6\textwidth]{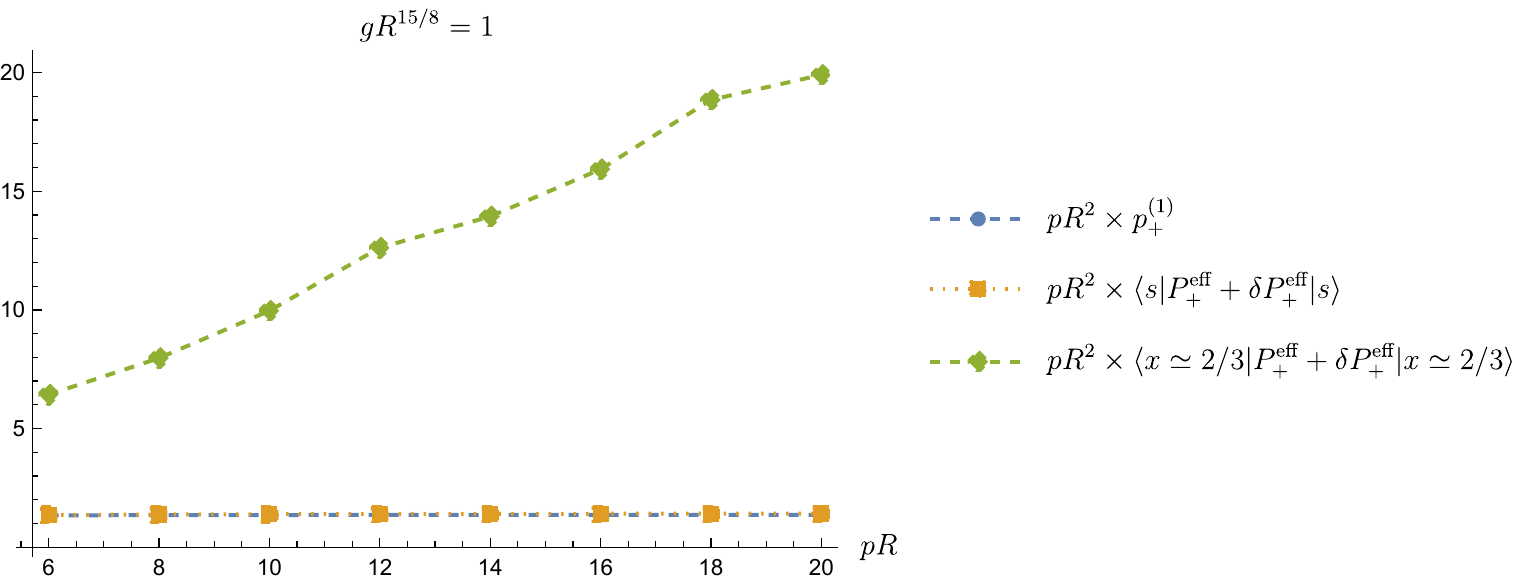}

\caption{$p_+^{(1)}$ is the first eigenvalue of $P_+^\textrm{eff}+\delta \Ppeff$, and the 2-particle state $|s\rangle$ is defined by $|s\rangle = \sum_x \sin (3\pi(x-1/2)/2)|x\rangle$, where $x$ is the parton momentum fraction. The expectation value of $p \times (\Ppeff+\delta \Ppeff)$ in the DLCQ 2-particle basis state $|x=2/3\rangle$ is seen to rise with $p$ indicating that it does not belong to the low-energy theory, while the smooth $|s\rangle$ state has a finite expectation as $p \rightarrow \infty$.  The $|s\rangle$ state wavefunction also happens to be a decent approximation of the actual lowest state wavefunction in the 2-particle sector.}
\label{fig:HeffME}
\end{center}
\end{figure}

We shall restrict our attention to the three lowest dimension chiral states of LCT and check the $M^2_\sigma$ matrix in this subsector at three different values of the coupling.
These states are $|1\> = L_{-2}|0\>$, $|2\> \propto (6 L_{-4} - 10 L_{-2}^2)|0\>$, and $|3\> \propto (6 L_{-4}L_{-2} - 15 L_{-3}^2 + 10 L_{-2}^3)|0\>$.
In what follows, we show the corresponding $3\times3$ matrix in the $p\rightarrow \infty$ limit. We factor out the volume dependence $\frac{g f(g,0,R)}{2p}$ to show that the remaining piece $M^2_\sigma$ is independent of the volume $R$ (or equivalently, the $g$ value) that we used to compute $\Ppeff$:
 \begin{equation}
\begin{aligned}
M^2_\sigma &= \left(
\begin{array}{ccc}
 22.58 & 12.93 & 10.13 \\
 12.93 & 73.26 & 41.92 \\
 10.13 & 41.92 & 131.3 \\
\end{array}
\right),  \qquad (g R^{2-\Delta} \sim 0), 
\\ 
M^2_\sigma &= \left(
\begin{array}{ccc}
 22.55 & 12.94 & 10.30 \\
 12.94 & 73.17 & 42.52 \\
 10.30 & 42.52 & 133.6 \\
\end{array}
\right), \qquad (gR^{2-\Delta}=0.4670), 
\\
M^2_\sigma  &= \left(
\begin{array}{ccc}
 22.53 & 12.84 & 9.984 \\
 12.84 & 72.56 & 40.91 \\
 9.984 & 40.91 & 129.3 \\
\end{array}
\right), \qquad  (gR^{2-\Delta}=1.713) . 
\end{aligned}
\label{eq:pPlusLCTAtDifferentR}
\end{equation}
 The ``$g R^{2-\Delta}\sim 0$'' matrix is computed by series expanding the effective Hamiltonion (\ref{eq:HeffIntro}) to leading nonzero order in $g$, 
 and then fixing $f(g,0,R)$ by requiring the first bound state mass $m_1$ extrapolated to $p\rightarrow \infty$ agrees with the infinite volume result 
 \begin{equation}
 m_{R=\infty} = 4.404921 \left(\frac{g}{2\pi}\right)^{\frac{8}{15}}.
 \label{eq:m1Infty}
 \end{equation} 
For $g R^{2-\Delta} =0.4670$ and $g R^{2-\Delta} =1.713$ cases, we follow the same procedure except that we use the finite values of $g$ indicated. 

Finally, we turn to the physical interpretation of the prefactor function $f$ in (\ref{eq:BoostedVolumeDependenceHeff}). One piece of intuition is that in lightcone quantization, the vacuum is not renormalized by interactions, so some structure of the interacting vacuum must be put in by hand.  In particular, primary operators such as $\sigma$ have vanishing vevs in the original CFT vacuum.  Instead, the interacting vacuum in the equal-time description develops a component of the Ramond vacuum, and the vev of $\sigma$ is one measure of how large this component is.  In Fig. \ref{fig:SigmaVevHeff}, we demonstrate that this intuition is remarkably precise: the prefactor function $f$ is exactly the volume-dependent vev of $\sigma$.  To make this comparison, we compute the mass of the lightest eigenstate of $\Ppeff$ at various different volumes (or equivalently, different couplings), and show the ratio of this mass to its infinite volume limit (\ref{eq:m1Infty}).  To obtain the volume-dependent $\< \sigma(R)\>$, we took the vacuum energy $E_{\rm vac}(R)$ from \cite{Klassen:1990dx} calculated using the thermodynamic Bethe Ansatz,\footnote{In terms of their scaling function $\tilde{c}(r)$, the vacuum energy $E_{\rm vac}(R)$ is $E_{\rm vac}(R) = 2 \pi R \Lambda g^{16/15}  - \frac{ \tilde{c}(2\pi R \cdot m_{R=\infty})}{12 R}$, where $ 2\pi \Lambda = - 1.06$.  At large $g$, the vev is $\<\sigma(\infty)\> = -\frac{16}{15} 2 \pi \Lambda g^{1/15}$. } and obtain $\< \sigma\>$ from $\<\sigma\> =- \frac{d}{dg} \frac{E_{\rm vac}(R,g)}{R}$. The agreement is excellent.

\begin{figure}
\begin{center}
 	 \includegraphics[width=0.6\linewidth]{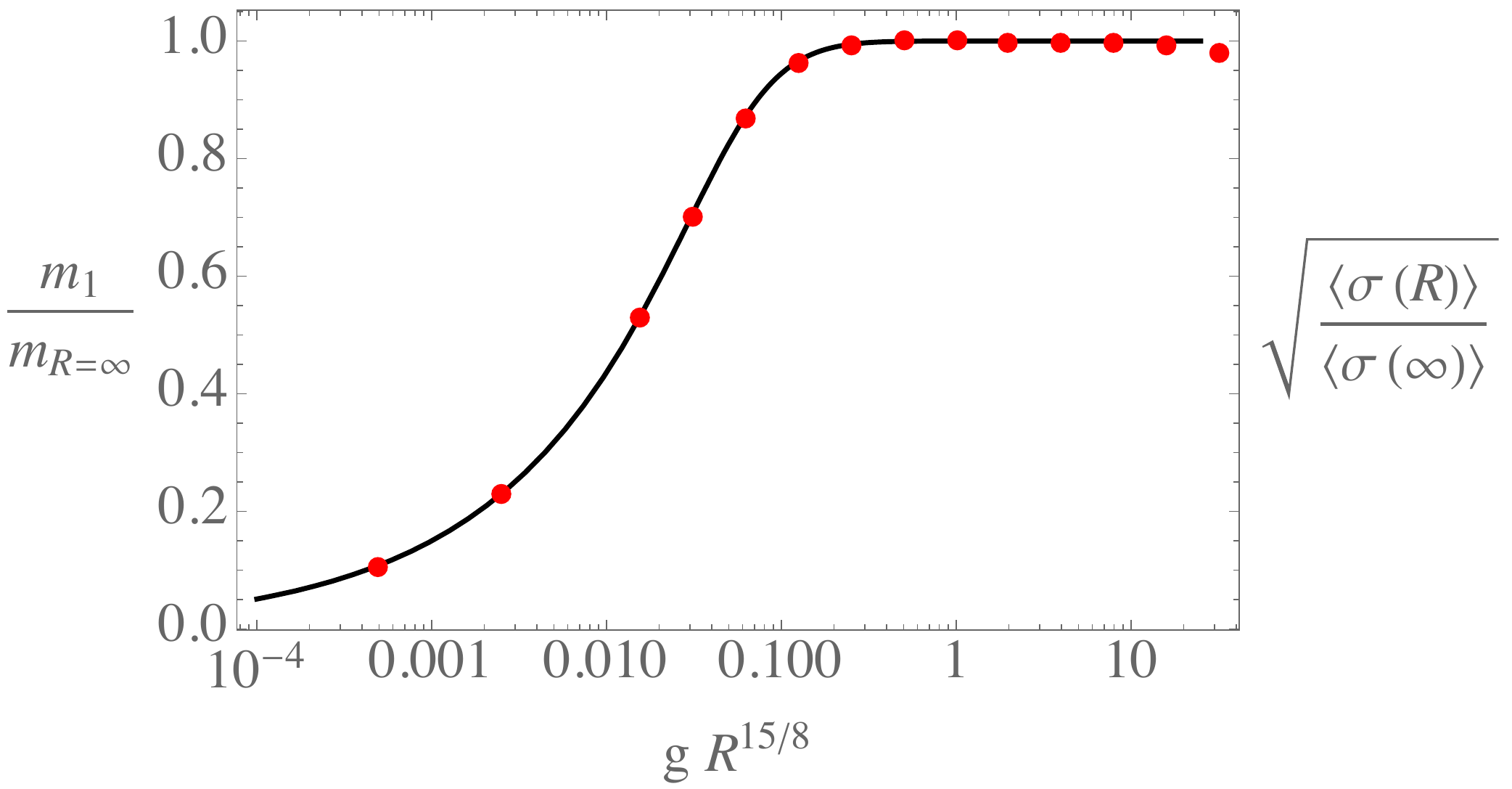}
	 \end{center}
	 \caption{Volume-dependent mass scale compared with the volume-dependent vev of $\sigma$.  {\it (Red dots)}: Ratio of lightest eigenvalue $m_1$ at finite volume $R$ to its value $m_{R=\infty}$ at infinite volume; this ratio is equivalent to $\sqrt{\frac{f(g,0,R)}{f(g,0,\infty)}}$ in terms of the prefactor $f$ in (\ref{eq:BoostedVolumeDependenceHeff}). {\it (Black, solid)}: Square root of the ratio of the vev $\<\sigma(R)\>$ at finite $R$ to its infinite $R$ value $\< \sigma(\infty)\>$, obtained from integrability results in \cite{Klassen:1990dx}.}
	 \label{fig:SigmaVevHeff}
	 \end{figure}

\subsection{Low Temperature Phase (\texorpdfstring{$\eta <0$}{eta<0})}
\label{sec:lowTemperaturePhase}

In this subsection we discuss the non-integrable regime where both $\epsilon$ and $\sigma$ are turned on, and $\eta <0$ so we are in the low temperature phase. We take the IFT Hamiltonian (\ref{eq:Hamiltonian}), compute the LC effective Hamiltonian to subleading order using (\ref{eq:EffectiveHamiltonian}) and (\ref{eq:EffectiveHamiltonianCorrection}), and use the Ward identity to determine the vacuum energy according to Appendix. \ref{app:Vac}. We compute the spectrum and Zamolodchikov $c$-funtion and compare them with the original Hamiltonian. To fix the vacuum energy, we require that the Ward Identity between $T_{+-}$ and $T_{--}$ from the original Hamiltonian holds for the lowest bound state. We show the ratio between the second and first bound state masses as a volume-independent observable in Figure \ref{fig:spectrumEtaMinus1}, and we show the Zamolodchikov $c$-function of the lowest bound state in Figure \ref{fig:cfuncEtaMinus1}. In Figure \ref{fig:spectrumEtaMinus1}, we see that the agreement between the original Hamiltonian $P_+$ and the effective Hamiltionian $P_+^{\rm eff}$ is fairly good, within about $\sim 10\%$ over the range of values shown, and improves significantly if we also include the leading $1/p$ correction, $\delta P_+^{\rm eff}$, to the effective Hamiltonian. In both cases, extrapolation to $p\rightarrow \infty$ gives a much better agreement.  
In Figure \ref{fig:cfuncEtaMinus1}, we see again that the $T_{--}$ acting on the interacting vacuum $|\Omega\>$ for $P_+$ agrees with $T_{--}^{ll}$ acting on the free vacuum for $\Ppeff$ at the percent level, displaying evidence of the `nonrenormalization' of $T_{--}$ (discussed in (\ref{eq:TmmNonRenorm})). By construction, $c_1$ measured by the original $P_+$ using the interacting vacuum $\Omega$ satisfies the Ward identity exactly since this condition is used to fix the effective vacuum energy in the boosted frame, and we see in the right panel that the Ward identity  holds for $c_2$ as well, giving a nontrivial check of the boosted Hamiltonian.
\begin{figure}[t!]
\begin{center}
\includegraphics[width=0.45\textwidth]{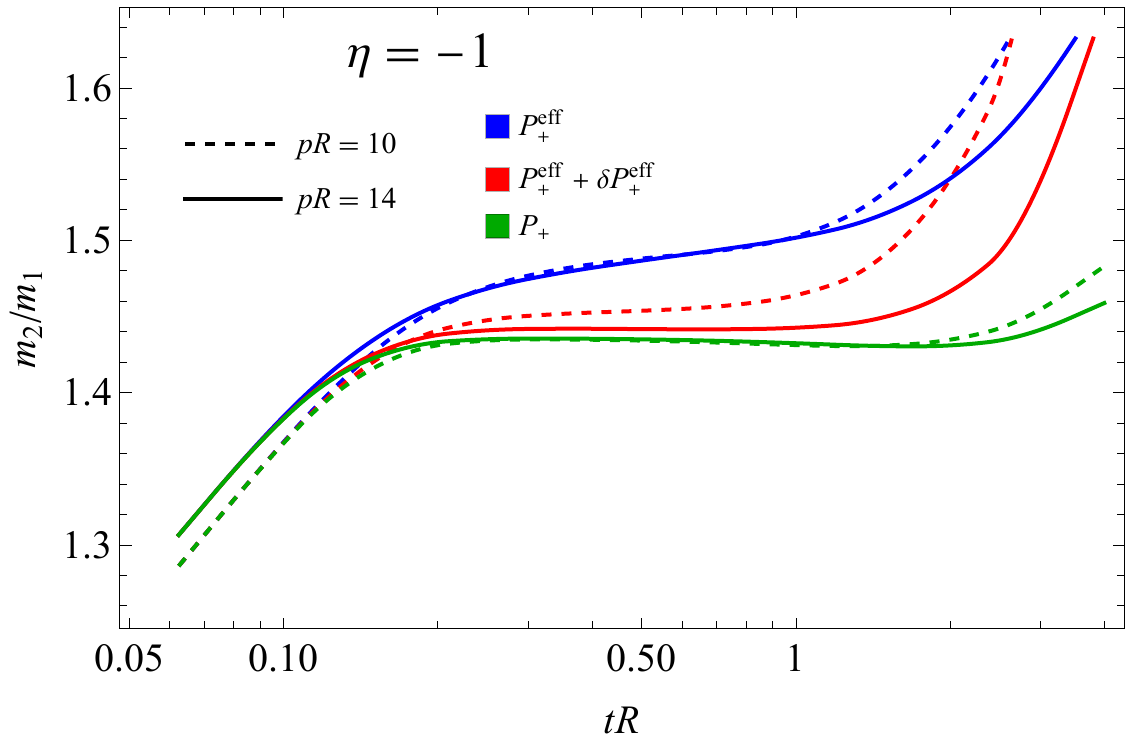}
\includegraphics[width=0.45\textwidth]{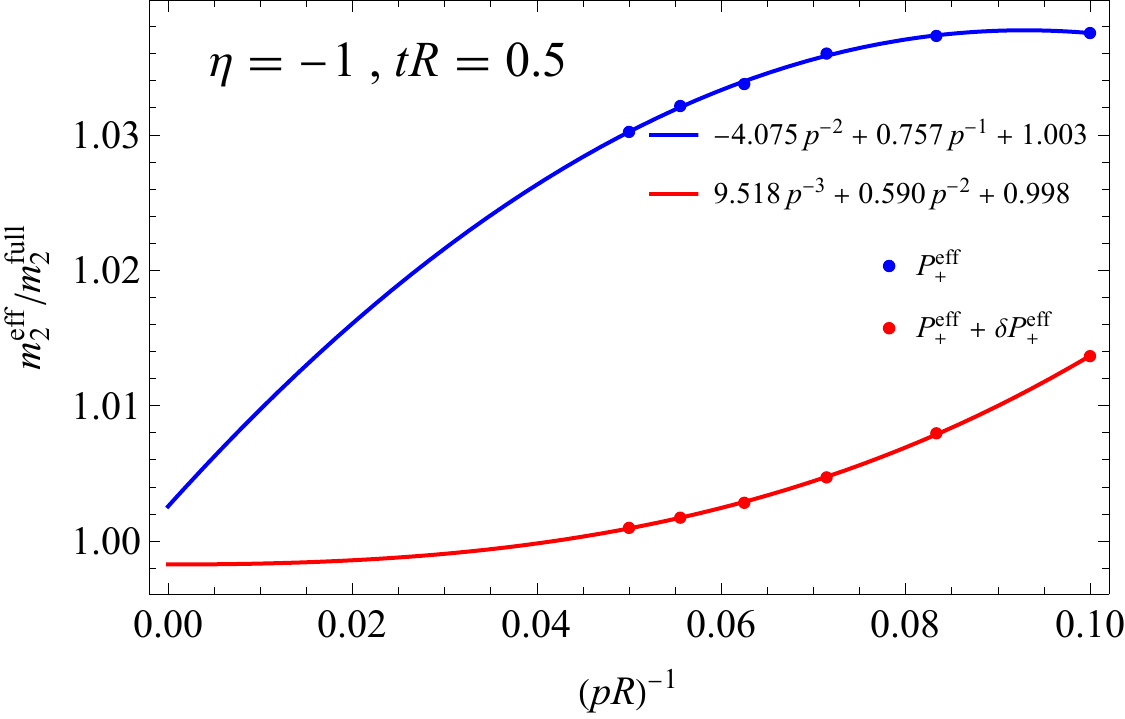}
\caption{
{\bf Left panel:} Second eigenvalue of $\Ppeff$ (or $\Ppeff + \delta \Ppeff$) and the original Hamiltonian $P_+$ for $\eta=-1$. We use $\Lambda R=40$ and compare $pR=10$ (dashed line) and $pR=14$ (solid line).  $\Ppeff$ and the original $P_+$ show better agreement and the measurements have weaker volume dependence at larger $p$.
{\bf Right panel:}
The convergence of the effective Hamiltonian at $tR=1/2$ ($t$ is defined in (\ref{dimensionlessVolume}) ). The blue and red points represent the ratio between $m_2$ measured from the effective Hamiltonian and the original Hamiltonian, for $\Ppeff$ and $\Ppeff+\delta \Ppeff$, respectively. The $\Ppeff$ data and $\Ppeff+\delta \Ppeff$ fit well to polynomial models starting at $p^{-1}$ and $p^{-2}$, respectively. Both measurements converge to $1$ at $p\rightarrow \infty$. 
}
\label{fig:spectrumEtaMinus1}
\end{center}
\end{figure}

\begin{figure}[h!]
\begin{center}
\includegraphics[width=0.45\textwidth]{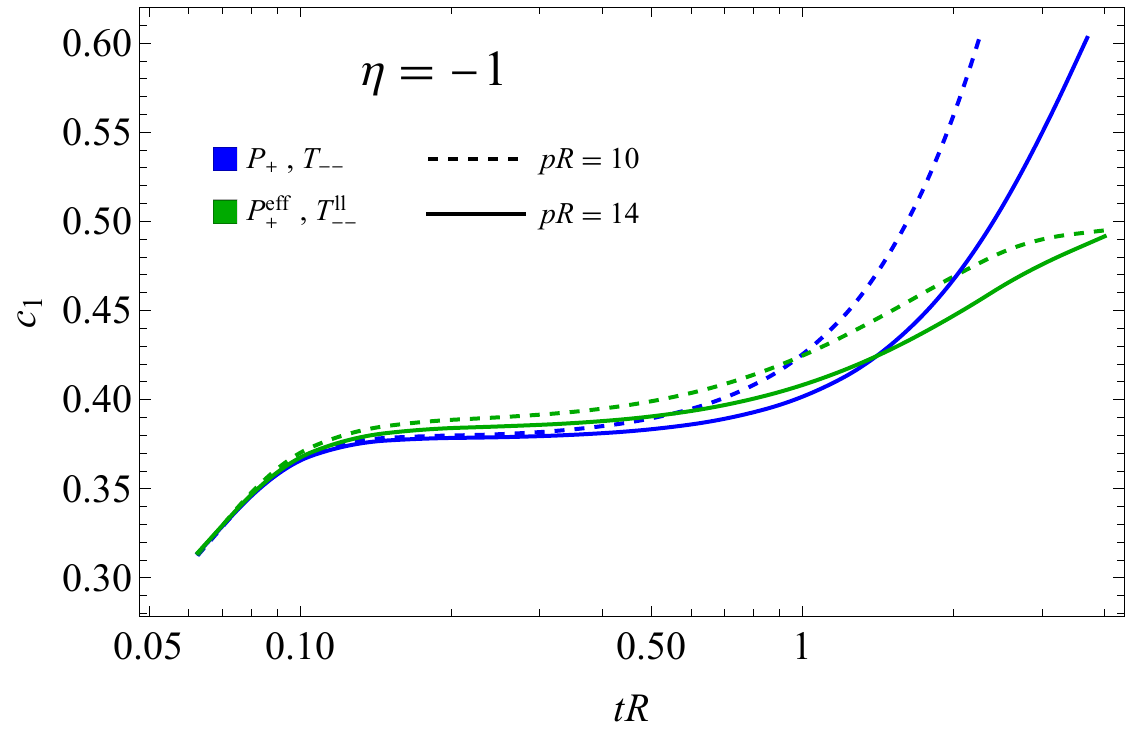}
\includegraphics[width=0.45\textwidth]{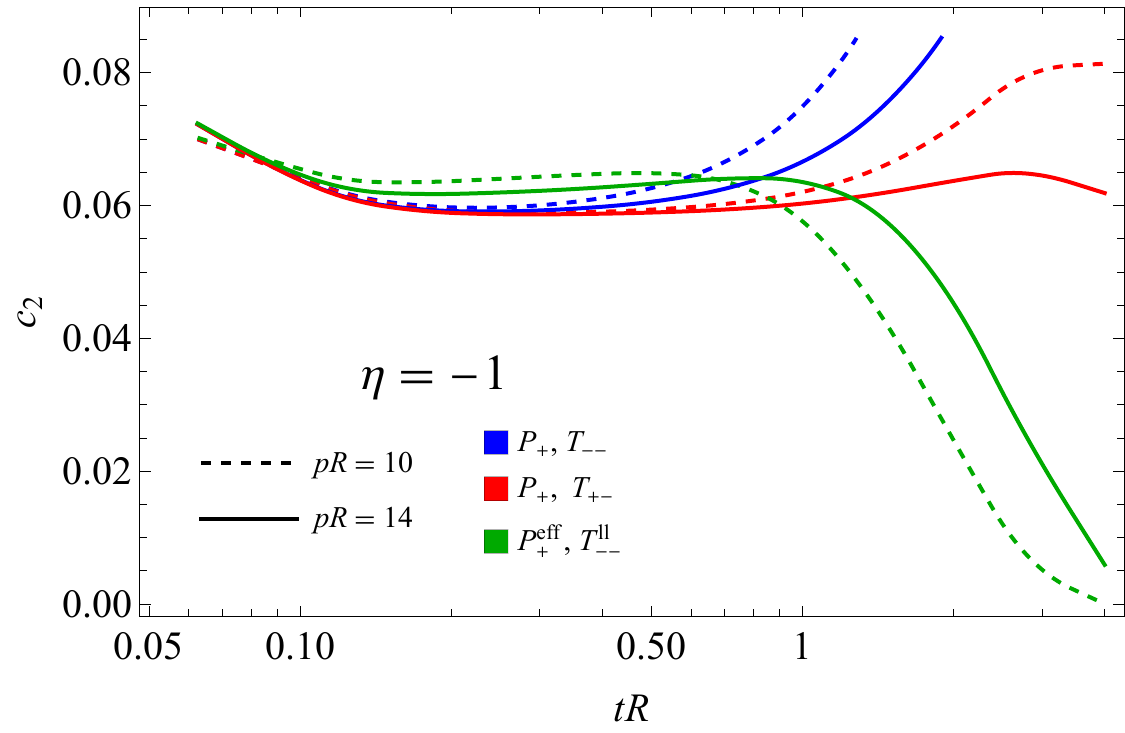}
\caption{ 
Bound state contributions to the Zamolodchikov $c$-function at $\eta = -1$. For the original $P_+$ we compute it using $T_{--}$ (blue line) and $T_{+-}$ (red line) and we use the interacting vacuum $|\Omega\>$. For $\Ppeff$ we us the Fock vacuum (green line, including the correction derived in section \ref{sec:effectiveHamiltonainCorrection}).  We use $\Lambda R=40$ and compare $pR=10$ (dashed line) and $pR=14$ (solid line). 
}
\label{fig:cfuncEtaMinus1}
\end{center}
\end{figure}

In the low temperature phase the LC effective Hamiltonian takes the form
\begin{equation}
\Ppeff =  m^2 \int dx^- \psi \frac{1}{2i\partial_-}\psi + \frac{g \langle\sigma\rangle}{2p} M^2_{\sigma} 
+  \CO(\frac{1}{p^2}).
\tag{\ref{eq:BoostedVolumeDependenceHeffIntro}}
\end{equation}
with the only dependence on volume contained in $\langle\sigma\rangle$. 
Thus, the drift in the measurements in both Fig \ref{fig:spectrumEtaMinus1} and Fig \ref{fig:cfuncEtaMinus1} at small volumes, $tR\lesssim .1$, can be understood as the $\sigma$ vev changing from its large volume value. The $c$-function measured at small volume is still physical, but agrees with a different $\eta$ measured at large volume (with the sigma vev determining the conversion to the right $\eta$). We will see in the next paragraph that the factor $\langle\sigma\rangle$ can be computed in certain regimes, giving us more evidence of the conjectured form (\ref{eq:BoostedVolumeDependenceHeffIntro}).

In Figure \ref{fig:pltSmallVolumeTCSAEtaInfPUpTo46Fit} we go to the regime $g /|m|^{\frac{15}{8}} \ll 1$. In this limit, fermions are confined and form an infinite family of approximate two fermion bound states starting from the two particle threshold. The bound state spectrum has been computed analytically \cite{PhysRevD.18.1259,Fonseca:2006au}.
\begin{equation}\label{analyticalSpectrumAnalytic}
m_n = 2m + (2\bar{\sigma} h)^{\frac{2}{3}} z_n \qquad (h\rightarrow 0)
\end{equation}
where $h = \frac{g}{2\pi}$, $\bar{\sigma} = |m|^{\frac{1}{8}} 2^{\frac{1}{12}}e^{-\frac{3}{2}\zeta'(-1)} = 1.35783834170660 |m|^{\frac{1}{8}}$ is the vacuum expectation of $\sigma$ at $h\rightarrow 0$, and $z_n$ is the $n$-th zero of Airy function ${\rm Ai}(-z)$. 
On the numerical TCSA side, we take $m R\ll 1$ and compute $\Ppeff$ perturbatively to leading order in $m$ and $g$. The analytical result (\ref{analyticalSpectrumAnalytic}) is in infinite volume and the numerical result $\Ppeff$ is at small volume.
The conjectured form (\ref{eq:BoostedVolumeDependenceHeffIntro}) suggests that at large momentum, the physical spectrum can be compared between different volumes. Due to the dependence on the order parameter $\langle \sigma \rangle$, there is a conversion factor between large volume and small volume coupling constants, $\frac{\langle\sigma(R\rightarrow \infty)\rangle}{\langle\sigma(R\rightarrow 0)\rangle}$. The large volume order parameter $\langle\sigma(R\rightarrow \infty)\rangle$ is the same as $\bar{\sigma}$, and the small volume order parameter $\langle\sigma(R\rightarrow 0)\rangle$ is computed perturbatively as $-\frac{d}{dg}E_{\rm vac} = \frac{4 \Gamma \left(\frac{1}{16}\right)^2 \Gamma \left(\frac{15}{8}\right)}{7 \Gamma \left(\frac{1}{8}\right) \Gamma \left(\frac{15}{16}\right)^2} g = 16.019 g$, see (\ref{perturbativeEvac}). The plot shows that at large momentum, the physical observables agree between infinite volume and the perturbative volume, where the computation is greatly simplified by the perturbative Hamiltonian. 
\begin{figure}[htbp]
\centering
\includegraphics[width=0.6\linewidth]{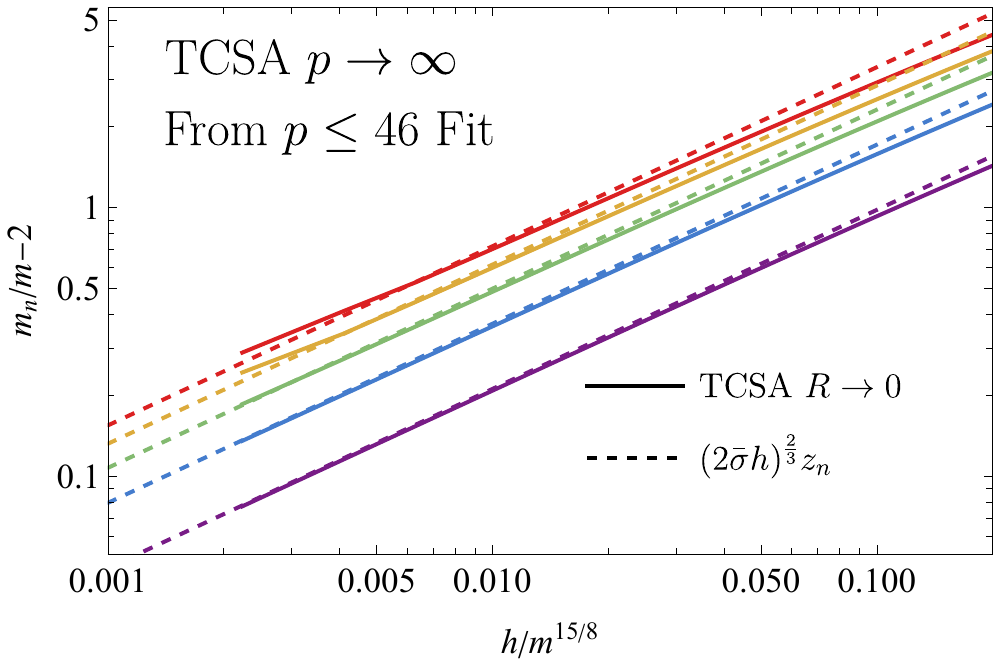}
\caption{\label{fig:pltSmallVolumeTCSAEtaInfPUpTo46Fit}
Matching of $P_+^{\rm eff}$ between small volume (solid lines) and infinite volume (dashed lines) at $h/m^{\frac{15}{8}}\ll 1 $. Lines of different colors represent different bound states. The infinite volume result is given by (\ref{analyticalSpectrumAnalytic}) and the small volume is computed from numerical $\Ppeff$ to leading order of $m$ and $g$, and extrapolated to $p R\rightarrow \infty$ using a fit linear in $\frac{1}{p}$. The two spectra match accurately for small but finite $h/m^{\frac{15}{8}}$, providing strong evidence for the conjectured form (\ref{eq:BoostedVolumeDependenceHeffIntro}). For smaller $h/m^{\frac{15}{8}}$ the numerical $\Ppeff$ breaks down due to worsening resolution at finite truncation. For larger $h/m^{\frac{15}{8}}$, (\ref{analyticalSpectrumAnalytic}) is not a good approximation of the binding energy. 
}
\end{figure}

\subsection{High Temperature Phase (\texorpdfstring{$\eta  >0$}{eta>0}) }
In this section we study the high temperature phase $\eta  >0$. Like in the previous section, we go to large momentum and compute $\Ppeff$ and its $1/p$ correction taking the chiral descendants of the NS vacuum as the low energy sector. 
We expect $\Ppeff$ to reproduce the low energy spectrum of $P_+$, and the $c$-function measured by $\< \Omega |T_{--}T_{--}| \Omega \>$ using the original $P_+$ to agree with $\< 0 |T_{--}^{ll}T_{--}^{ll}| 0 \>$ using $\Ppeff$.
In Figure \ref{fig:spectrumEtaPlus1Compare} we show the ratio of the first two bound state masses $m_2/m_1$. At intermediate volumes, $tR\sim 0.5$, and momentum $pR = 14$, $\Ppeff$ agrees with the original $P_+$ to a few percent,  with the error improved to less than a percent with the correction $\delta\Ppeff$. If we extrapolate to $pR \rightarrow \infty$, the agreement of both $\Ppeff$ and $\Ppeff+\delta\Ppeff$ with $P_+$ is much better. In Figure \ref{fig:cfuncEtaPlus1} we show the bound state contribution to the Zamolodchikov $c$-function, $c_1$ and $c_2$. While $c_1$ from  $\Ppeff$ agrees with the original $P_+$ to a few percent, $c_2$ has a large error. A possible reason for the large error in $c_2$ is that the wave function of the second bound state has to be orthogonal to the first one, and $c_2$ is small. A small error in $c_1$ can cause $c_2$ to change drastically.
\begin{figure}[h!]
\begin{center}
\includegraphics[width=0.45\textwidth]{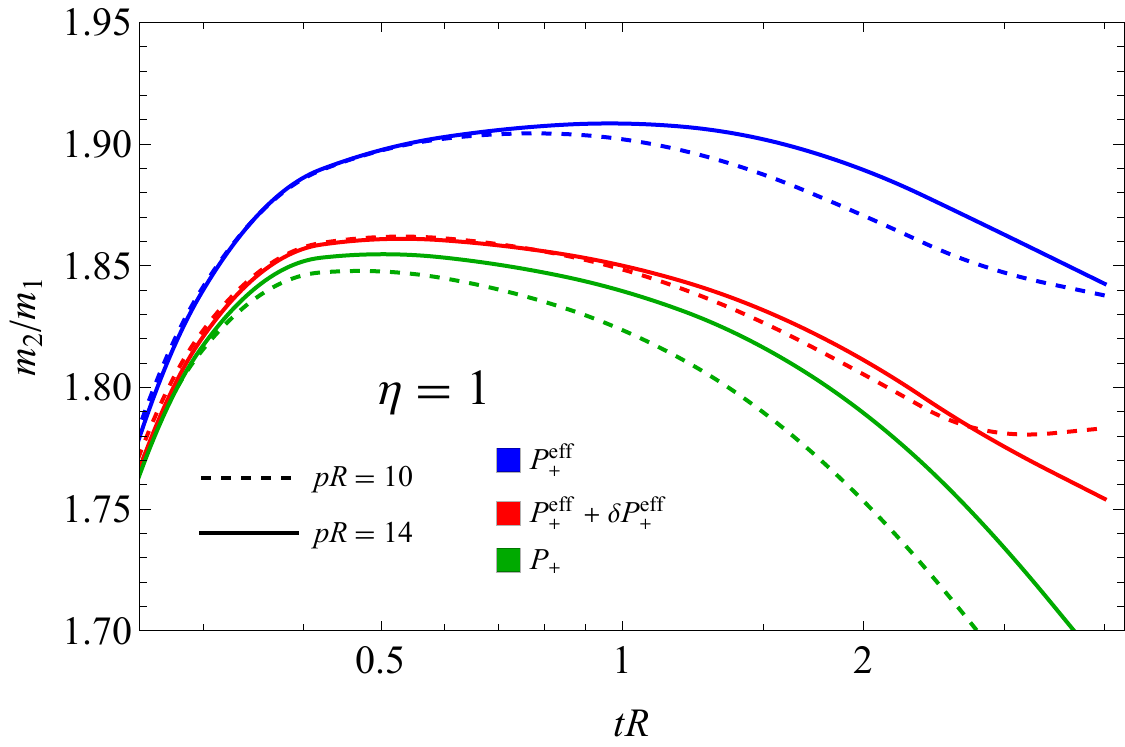}
\includegraphics[width=0.45\textwidth]{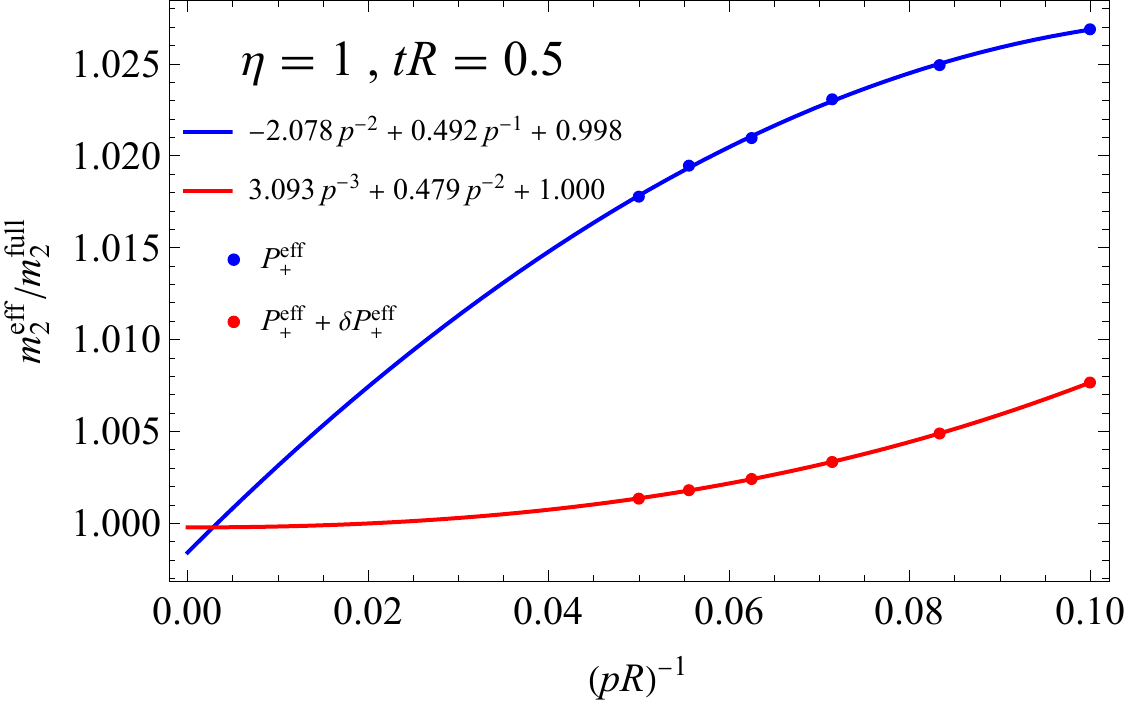}
\caption{
{\bf Left panel:} Second eigenvalue of $\Ppeff$ (or $\Ppeff+\delta \Ppeff$) and the original Hamiltonian $P_+$ for $\eta=+1$. We use $\Lambda R=40$ and compare $pR=10$ (dashed line) and $pR=14$ (solid line). 
{\bf Right panel:}
The convergence of the effective Hamiltonian at $tR=1/2$. The blue and red points represent the ratio between $m_2$ measured from the effective Hamiltonian and the original Hamiltonian, for $\Ppeff$ and $\Ppeff+\delta \Ppeff$, respectively. The $\Ppeff$ data and $\Ppeff+\delta \Ppeff$ fit well to polynomial models starting at $p^{-1}$ and $p^{-2}$, respectively. Both result converge to $1$ at $p\rightarrow \infty$. 
}
\label{fig:spectrumEtaPlus1Compare}
\end{center}
\end{figure}
\begin{figure}[h!]
\begin{center}
\includegraphics[width=0.45\textwidth]{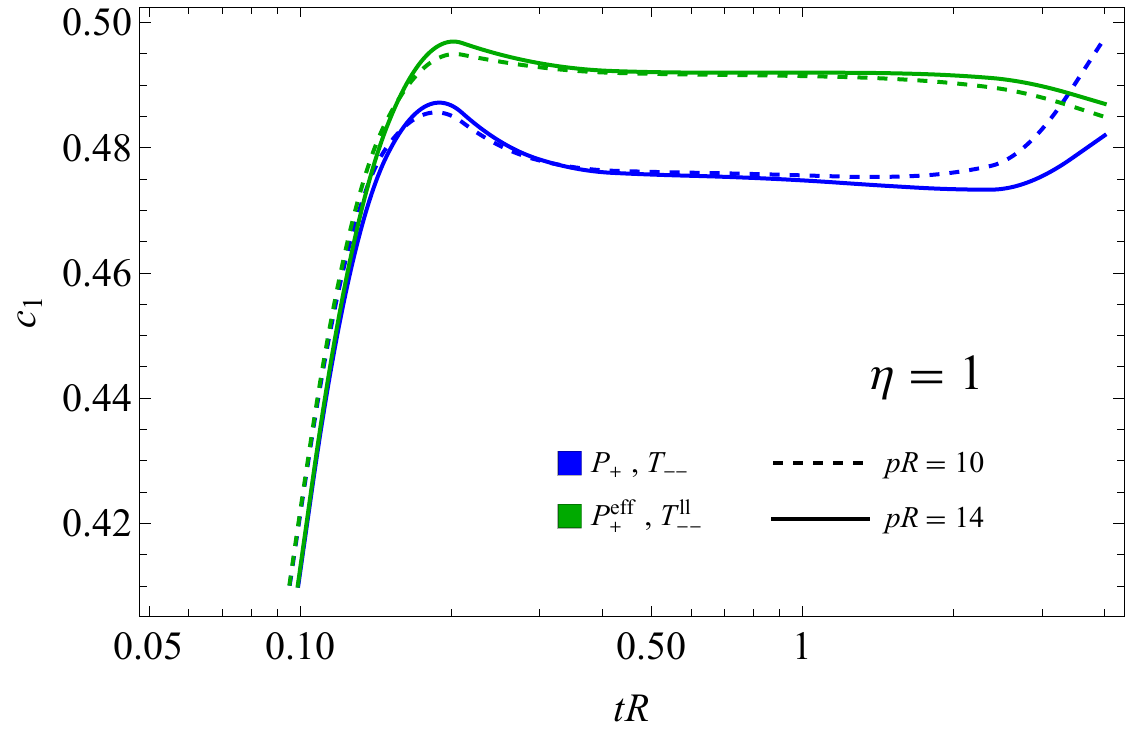} 
\includegraphics[width=0.45\textwidth]{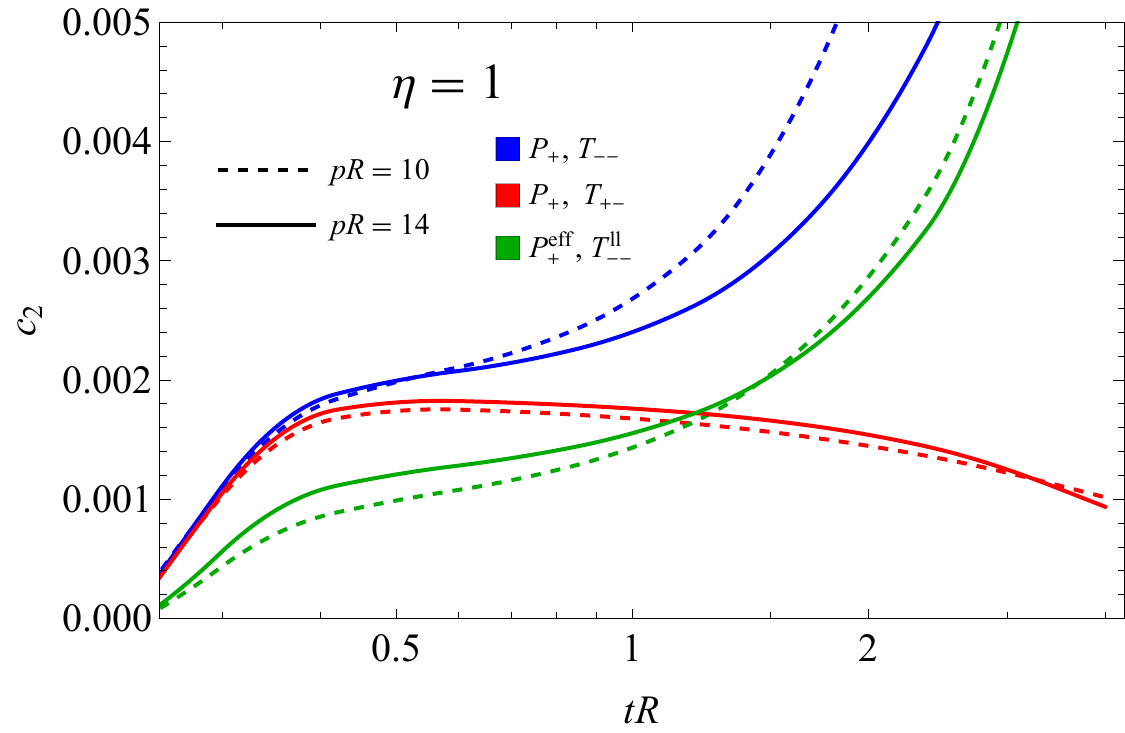} 
\caption{ 
Bound state contributions to the Zamolodchikov $c$-function at $\eta = +1$. For the original $P_+$ we compute it using $T_{--}$ (blue line) and $T_{+-}$ (red line) and we use the interacting vacuum $|\Omega\>$. For $\Ppeff$ (green line, including the correction derived in section \ref{sec:effectiveHamiltonainCorrection}).  We use $\Lambda R=40$ and compare $pR=10$ (dashed line) and $pR=14$ (solid line). 
}
\label{fig:cfuncEtaPlus1}
\end{center}
\end{figure}
The fact that $m_2$ and $c_1$ from $\Ppeff$ agree with original $P_+$, and that 
the extrapolation in Figure \ref{fig:spectrumEtaPlus1Compare} to large $p$ appears to converge to 1, provides a nontrivial check that our procedure constructs the correct effective Hamiltonian in the high temperature phase. The numerical results seem to suggest that at finite $p$ the broken phase $\Ppeff$ converges better than the unbroken phase $\Ppeff$. We expect that the convergence will become more of an issue for large positive $\eta$ because in the unbroken phase the odd number of fermion states are in the low energy spectrum, and their wave functions receive large contributions from the R sector in the original $P_+$ formulation. 
We have assumed that the interacting low energy sector contains only chiral descendants of the $NS$ vacuum. For the high temperature phase, in the limit $g\rightarrow 0$ this is not true because we have separate low energy sectors for R and NS states, which represent odd and even number of free massive fermions, respectively. For finite but small $g$ the separation between the interacting R and NS sectors hence may be small, requiring quite a large $p$ for $\Ppeff$ to converge. 

To address convergence with $p$, we can choose instead to keep the chiral descendants of the Ramond sector in the low energy sector, and compute a new $\Ppeff$ using the same procedure, but for this larger sector. In this alternative setup the $p=0$ low energy sector is also larger, with two states: The NS vacuum and R vacuum.   The LC vacuum is 
thus a certain linear combination of these two vacua
\begin{equation}
|\Omega\rangle_{\rm LC} = \cos \theta |{\rm vac}\rangle_{\rm NS} + \sin \theta |{\rm vac}\rangle_{\rm R}
\end{equation}
where the mixing angle $\theta$ is a function of $\eta$. 
We test this setup in Figure \ref{fig:pltCFuncHeffRamond}, where we compute the $c$-function from $\Ppeff$ including the Ramond sector and compare with the original $P_+$ result. The $c$-function from $\Ppeff$ using the trivial NS vacuum is incorrect, while the spectrum agrees well with the original $P_+$, suggesting that the mixing angle $\theta$ is important to obtain the correct $c$-function as $T_{--}$ acts on this mixed LC vacuum. 
We would like to see if there exists a mixing angle $\theta$ which produces the correct $c$-function, and we determine this angle numerically by requiring $c_1$ from $\Ppeff$ to agree with the original $P_+$. With this procedure we can reproduce the whole $c$-function curve well using $\Ppeff$ with this one additional input. Finally, we compare the result at $\eta=1$ with previous results without the Ramond sector, shown in Figure \ref{fig:pltCFuncHeffRamondEta1Compare}. The $\Ppeff$ has better agreement with the original $P_+$ if chiral descendants of the Ramond vacuum are included in the low energy sector.
\begin{figure}[htbp]
\centering
\includegraphics[width=0.45\linewidth]{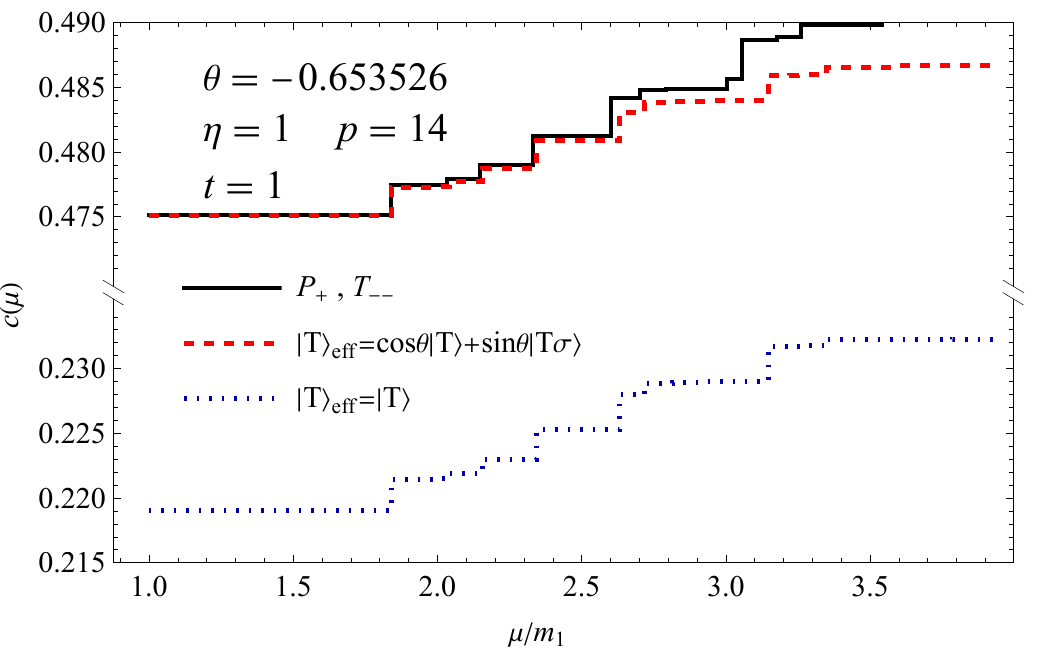}
\includegraphics[width=0.45\linewidth]{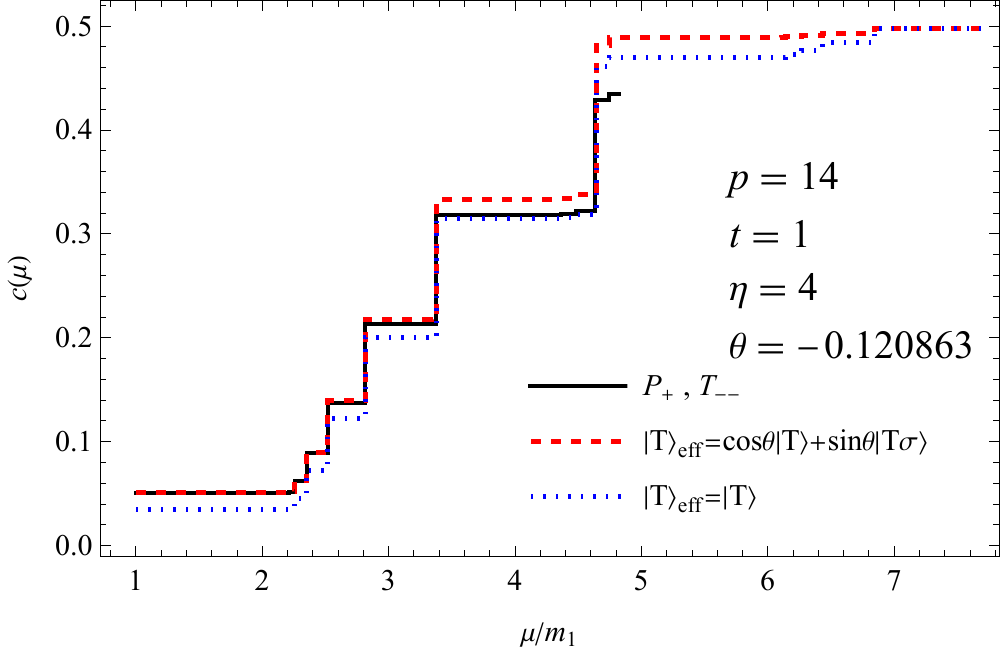}
\caption{\label{fig:pltCFuncHeffRamond}
Zamolodchikov $c$-function obtained from $\Ppeff+\delta \Ppeff$ with Ramond sector (red dashed line) and from the original $P_+$ (black solid line). 
The $c$-function is computed with the $T_{--}$ acting on a mixture of R- and NS- vacuum, where the mixing angle $\theta$ is chosen such that $c_1$ matches the original $P_+$ result.
The blue dotted line represents the same computation of $\Ppeff+\delta \Ppeff$ with the Ramond sector but ignoring the vacuum mixing, i.e. $\theta = 0$. We see that $\Ppeff$ with Ramond sector accurately reproduces the original $P_+$ results at both small and large $\eta$, and the vacuum mixing is important for small $\eta$.
}
\end{figure}
\begin{figure}[htbp]
\centering
\includegraphics[width=0.6\linewidth]{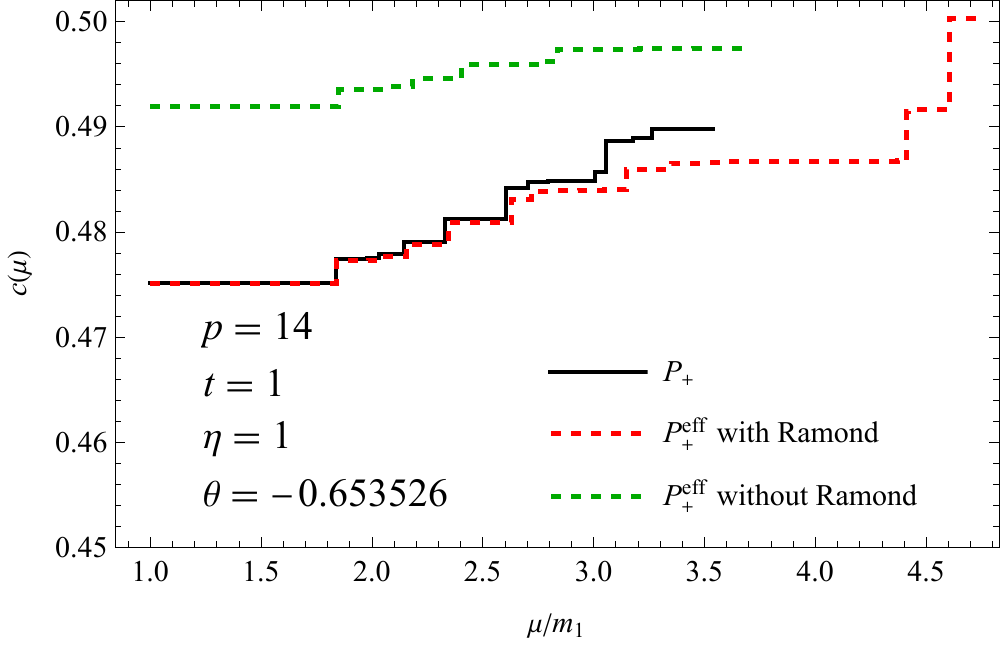}
\caption{\label{fig:pltCFuncHeffRamondEta1Compare}
The $c$-function at $\eta=1$. The red dashed line is obtained with $\Ppeff+\delta \Ppeff$ taking the chiral descendants of both NS vacuum and R vacuum as the low energy sector, the green dashed line is obtained with $\Ppeff+\delta \Ppeff$ without the Ramond sector, and the black solid line is the original $P_+$ result. The effective Hamiltonian that includes the Ramond sector has better agreement with the original Hamiltonian.
}
\end{figure}
 
\section{Relation to Dyson Series}
\label{sec:Dyson}

The expression (\ref{eq:EffectiveHamiltonian}) for the effective Hamiltonian $\Ppeff$ is nonperturbative, but it can be expanded perturbatively in the coupling.  An alternate, {\it perturbative}  expression for the effective Hamiltonian was formulated in \cite{Fitzpatrick:2018ttk} in terms of the Dyson series of the unitary evolution operator $U(t_1, t_2)$ in the interaction picture of QFT.  Here, we will clarify the relation between the perturbative expansion of  (\ref{eq:EffectiveHamiltonian}) and the formulation in \cite{Fitzpatrick:2018ttk}.   

First, we review the perturbative construction.  Roughly, the construction is to take the following limit of the connected\footnote{By ``connected'', we mean Feynman diagrams where every subgraph is connected to at least one external line, in other words there are no  bubbles.} part of $U$:
\begin{equation}
\< l | P_+^{\rm eff} | l'\> = \lim_{x^+ \rightarrow 0} \lim_{p_x \rightarrow \infty} i \partial_{x^+} \<l | U_{\rm conn}(x^+ ) | l'\>.
\label{eq:DysonDef1}
\end{equation}
We will see below that an equivalent way to write the connected component is to take
\begin{equation}
\< l | P_+^{\rm eff} | l'\> = \lim_{x^+ \rightarrow 0} \lim_{p_x \rightarrow \infty} \frac{i \partial_{x^+} \<l | U(x^+ ) | l'\>}{\<l | U(x^+ ) | l'\>},
\label{eq:DysonDef2}
\end{equation}
which has the advantage that it does not require a Lagrangian description so that one can remove `bubble' diagrams.

To make the definition above precise, we need to define the limits more carefully, which we will do below.  
At small coupling, it is possible to obtain an arbitrarily large hierarchy between the eigenstates of $P_+$ containing only left-moving particles, and all other states, since the former will have eigenvalues suppressed by powers of the coupling whereas the latter will have eigenvalues at least as large as the smallest possible right-moving momentum $\sim \CO(1/R)$.  For convenience, let $E_l$ and $E_h$ be lightcone  energies characteristic of these two sectors;
at small enough coupling we have $E_l \ll E_h$. The limit ``$x^+\rightarrow 0$'' should more accurately be understood now as $E_h^{-1} \ll x^+ \ll E_l^{-1}$; moreover, we will see that rather than picking a single value of $x^+$, it must be averaged over a small window.  Define $\Pi_{E_l}$ to be the projector onto the light eigenstates:
\begin{equation}
\Pi_{E_l} \equiv \sum_{E_n \textrm{ light e-states}} |\hat{n} \> \< \hat{n} | .
\end{equation}
From (\ref{eq:Wdef}) and (\ref{eq:WSZ}), we have
\begin{equation}
\< l |  Z^{-1} |l'\>  = \< l | \Pi_{E_l} | l'\>,
\end{equation}
or in other words, $Z^{-1} = \Pi_{E_l}$ within the subspace spanned by the $|l\>$ basis states.

The next step is to consider $i \partial_{x^+} U$.  It is slightly more aesthetically pleasing to consider the following derivative, which is manifestly Hermitian at finite $x^+$:
\begin{equation}
\frac{1}{2} i \partial_{x^+} U(x^+,-x^+) = \frac{1}{2} e^{i P_{+,0} x^+} \left( (P_{+} - P_{+,0})e^{-2 i P_+ x^+} + e^{-2 i P_+ x^+} (P_{+} - P_{+,0})\right) e^{i P_{+,0} x^+}.
\end{equation}
We will evaluate matrix elements of this operator between two basis states $\< l | \dots | l'\>$ by inserting a complete set of eigenstates of $P_+$:
\begin{equation}
\begin{aligned}
&\< l | \frac{1}{2} i \partial_{x^+} U(x^+,-x^+) | l'\> = \\
& \frac{1}{2} \< l | e^{i P_{+,0} x^+} \left( ( P_{+} - P_{+,0}) e^{ -2 i P_+ x^+} | \hat{n} \> \< \hat{n} | +  | \hat{n} \> \< \hat{n} |e^{ -2 i P_+ x^+} ( P_{+} - P_{+,0})  \right) e^{i P_{+,0} x^+}| l'\> 
\end{aligned}
\end{equation}
Now, the key point is that the factor $e^{-2 i P_+ x^+}$ acts like a projector onto the light eigenstates of $P_+$. The reason for this is that for the heavy eigenstates, it is oscillating rapidly, and they get cancelled out, whereas for the light eigenstates we have $P_+ x^+ \approx 0$.  This is the reason for the requirement we mentioned above that we average over $x^+$. In Fig. \ref{fig:HfromAverages}, we show this explicitly for the case of the Ising model deformed by a mass term.  Similarly, we can approximate $e^{i P_{+,0} x^+} \approx 1$.  Therefore, we get
\begin{equation}
\begin{aligned}
\< l | \frac{1}{2} i \partial_{x^+} U(x^+,-x^+) | l'\>_{\rm avg} &\approx  \< l | \left( \sum_{E_n \atop  \textrm{ light  evals}} | \hat{n}\> E_n \< \hat{n}|- \frac{1}{2} ( P_{+,0} \Pi_{E_l} + \Pi_{E_l}P_{+,0})  \right) | l'\> \nn\\
 &=  \< l | \left(Z^{-\frac{1}{2}} P_+^{\rm eff} Z^{-\frac{1}{2}} - \frac{1}{2} (P_{+,0} Z^{-1} + Z^{-1} P_{+,0}) \right) | l'\>,
\end{aligned}
 \end{equation}
 where in the last line, we have used equations (\ref{eq:Wdef})-(\ref{eq:PpEffProj}), and the subscript ``$\textrm{avg}$'' indicates the time averaging we have discussed.

Next, let us justify the equivalence of (\ref{eq:DysonDef1}) and (\ref{eq:DysonDef2}) in the case of Lagrangian theories.  
For a single correlation function, the sum over all its diagrams is the product of two independent factors: a) the sum over all its connected diagrams, and b) the sum over all disconnected, or ``bubble'', diagrams:
\begin{equation}
\textrm{all diagrams}= \left[  \textrm{sum over connected diagrams} \right] \times \left[ \textrm{sum over bubble diagrams}\right].
\label{eq:AllVsConn}
\end{equation}
  The latter is just the contribution for the trivial vacuum-to-vacuum correlator, 
\begin{equation}
\lim_{x^+ \rightarrow \infty(1-i \epsilon)} \< 0 | U(x^+, -x^+) | 0 \>  = |\<0 | \Omega\>|^2 e^{-2 i E_\Omega x^+} \sim \left[ \textrm{sum over  bubble  diagrams}\right].
\end{equation}
As mentioned previously, we assume without loss of generality  that the vacuum energy $E_\Omega$ vanishes.  It may seem strange that we are invoking the limit $x^+\rightarrow \infty$ when our prescription has $x^+ \rightarrow 0$.  But the key point is that there is a hierarchy of energies, and we want $x^+$ to be much larger than $1/E_h$ (one over the high energy scale), and much shorter than $1/E_l$ (one over the low energy scale).  Because of (\ref{eq:AllVsConn}) and (\ref{eq:Zvac}), the effect of taking the connected piece of $U$ is to multiply by $Z_{p_x=0}$.
Using similar logic to the arguments above, we easily see that we can write $Z^{-1}$ as
\begin{equation}
Z^{-1} = \< l | \hat{n} \> \< \hat{n} | l'\> = \< l |U(x^+,-x^+) | l'\>_{\rm avg},
\end{equation} 
where as before, the average over $x^+$ is taken over  a regime where $E_h x^+ \gg 1 \gg E_l x^+$.  Therefore, we can actually compute  $Z^{-1}$ and $P_+^{\rm eff}$ using $U(x^+,-x^+)$. In Fig.\ \ref{fig:HfromAverages}, we check this numerically in the case of the Ising model with a mass deformation, where moreover one can see explicitly the need for performing an average over $x^+$ in order to extract $P_+^{\rm eff}$ and $Z^{-1}$.

 The last step is that we would like to commute the factors of $Z^{-1}$ past $P_{+,0}$ and $P_+^{\rm eff}$ and treat them as a constant in the denominator, as in (\ref{eq:DysonDef2}).   At least in the case of perturbative Lagrangian theories, we expect that in the limit of large $p$ and large volume, the matrix $Z$ in all momentum sectors simply becomes the contribution from disconnected diagrams, which is independent of the external state and therefore equal to $Z_{p_x=0}$ times the identity matrix.\footnote{One can check this explicitly in the case of a free scalar field deformed by a mass term.  The wavefunction for any multiparticle state is just a finite set of creation operators acting on the wavefunction for the ground state, which can be written in closed form in a free theory.  The overlap between, say, a one-particle state wavefunction in the massless and massive theory, can then be computed by performing a simple integral of a Gaussian times a polynomial, and by explicit computation one finds $|\< \Psi_p, m=0| \Psi_p, m\ne 0\>|^2 = |\< {\rm vac}, m=0 | {\rm vac}, m\ne 0\>|^2 \left( \frac{4 p \sqrt{m^2 + p^2}}{(p+\sqrt{m^2+p^2})^2}\right)$, which reduces to the universal factor  $|\< {\rm vac}, m=0 | {\rm vac}, m\ne 0\>|^2$ at $p=\infty$.  More generally, in a CFT deformed by a single relevant operator, by dimensional analysis large $p$ is related to small values of the relevant coupling, i.e., the undeformed theory. This argument assumes that the additional dependence in the wavefunction overlap arising from a high momentum mode is independent of the volume of space, which seems reasonable on physical grounds for a short wavelength excitation.   }   Assuming this is true, we finally obtain
\begin{equation}
\frac{\< l | \frac{1}{2} i \partial_{x^+} U(x^+,-x^+) | l'\>_{\rm avg}}{\<l | U(x^+ ,-x^+) | l'\>_{\rm avg}}  = \< l | (P_+^{\rm eff}- P_{+,0}  )| l'\>,
\end{equation}
so that the two constructions of $P_+^{\rm eff}$ agree perturbatively.

\begin{figure}[ht!]
\centering
\includegraphics[width=0.3\linewidth]{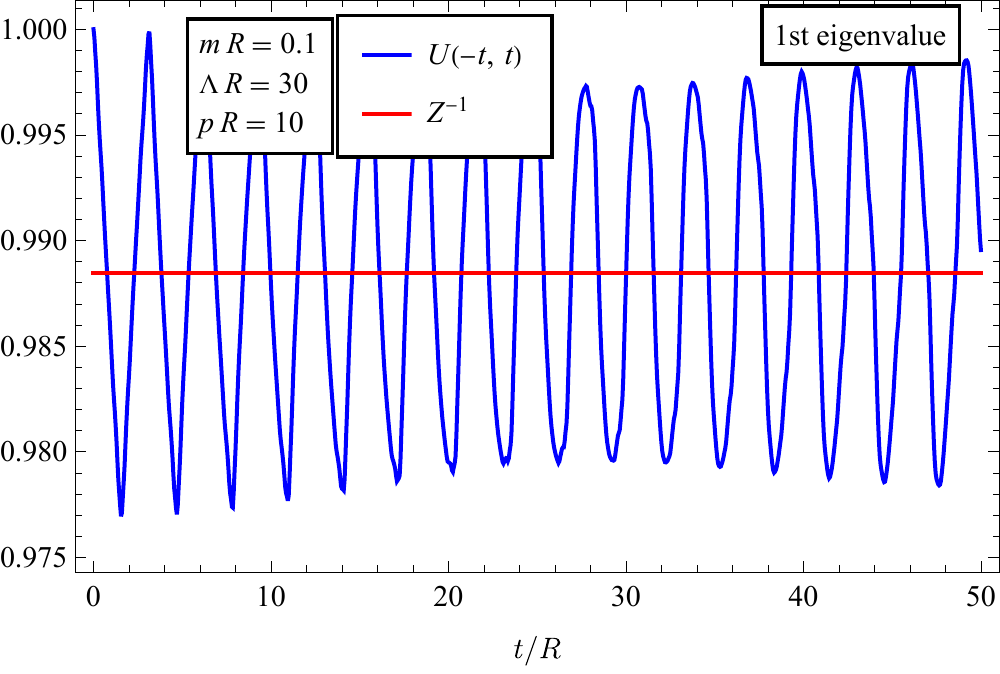}
\includegraphics[width=0.3\linewidth]{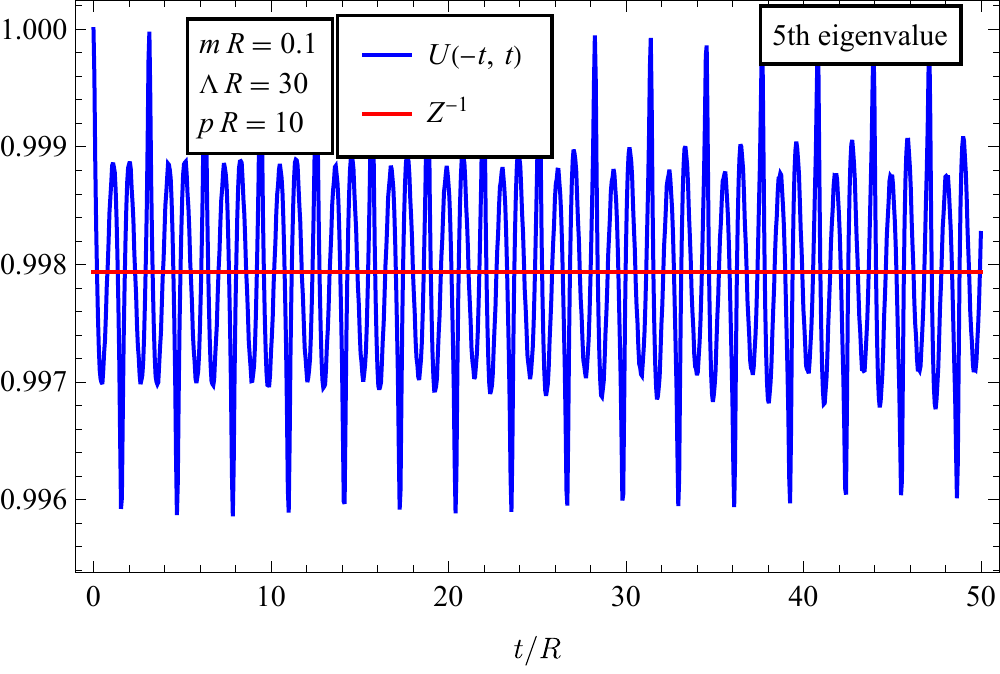}\\ 
\includegraphics[width=0.3\linewidth]{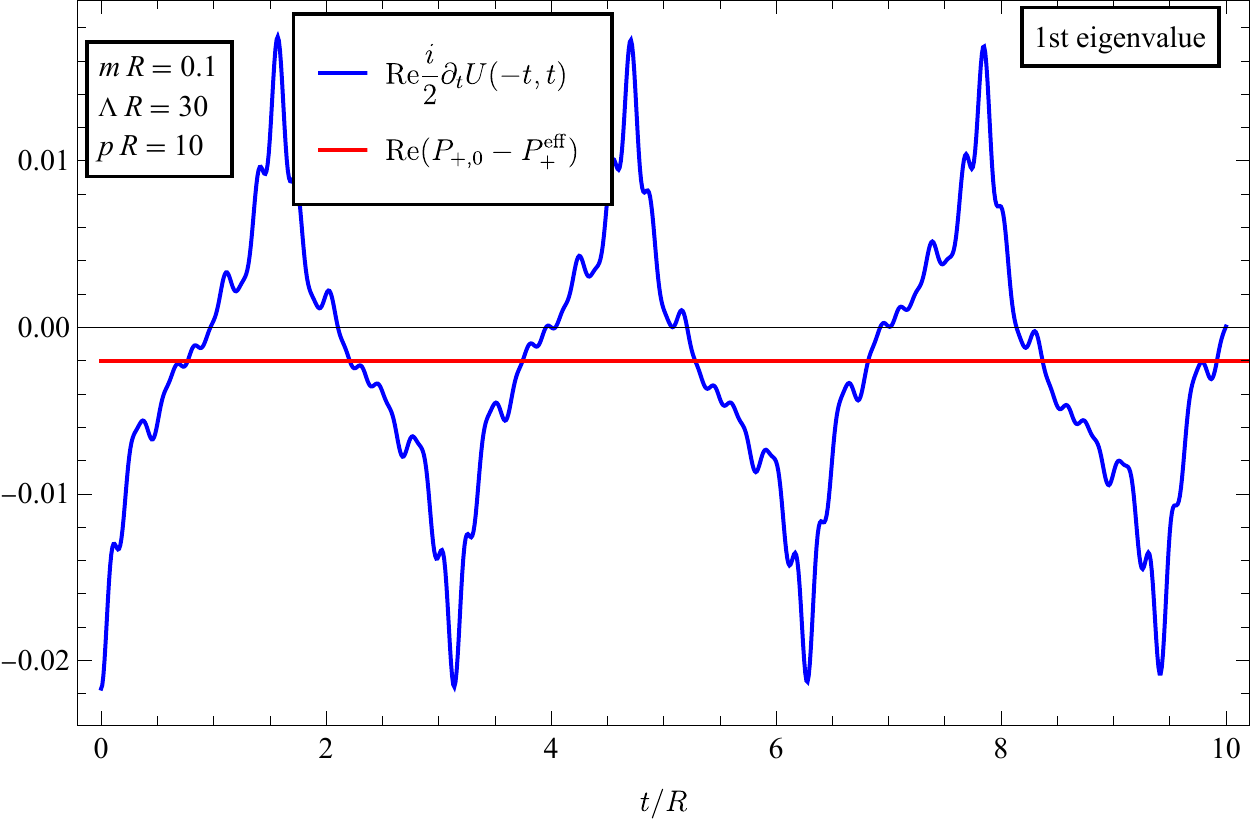}
\includegraphics[width=0.3\linewidth]{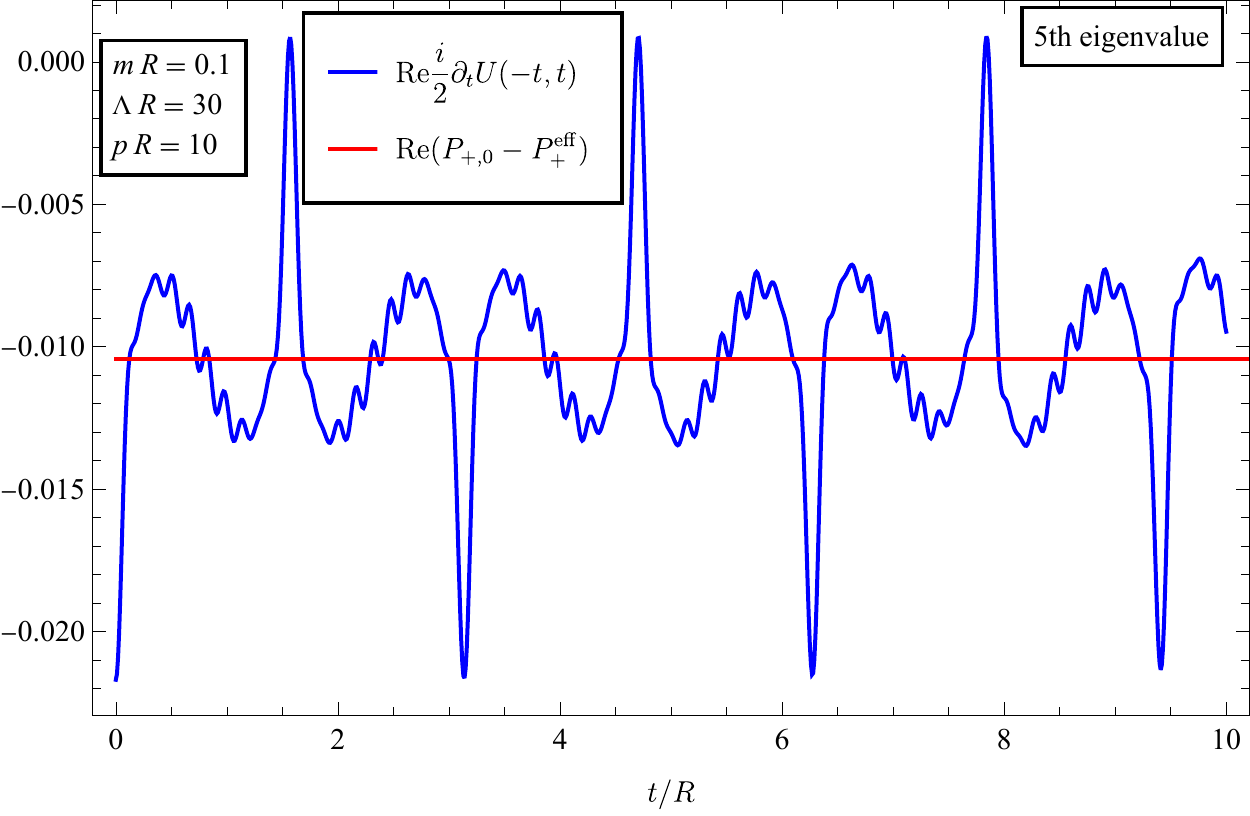}\\ 
\includegraphics[width=0.3\linewidth]{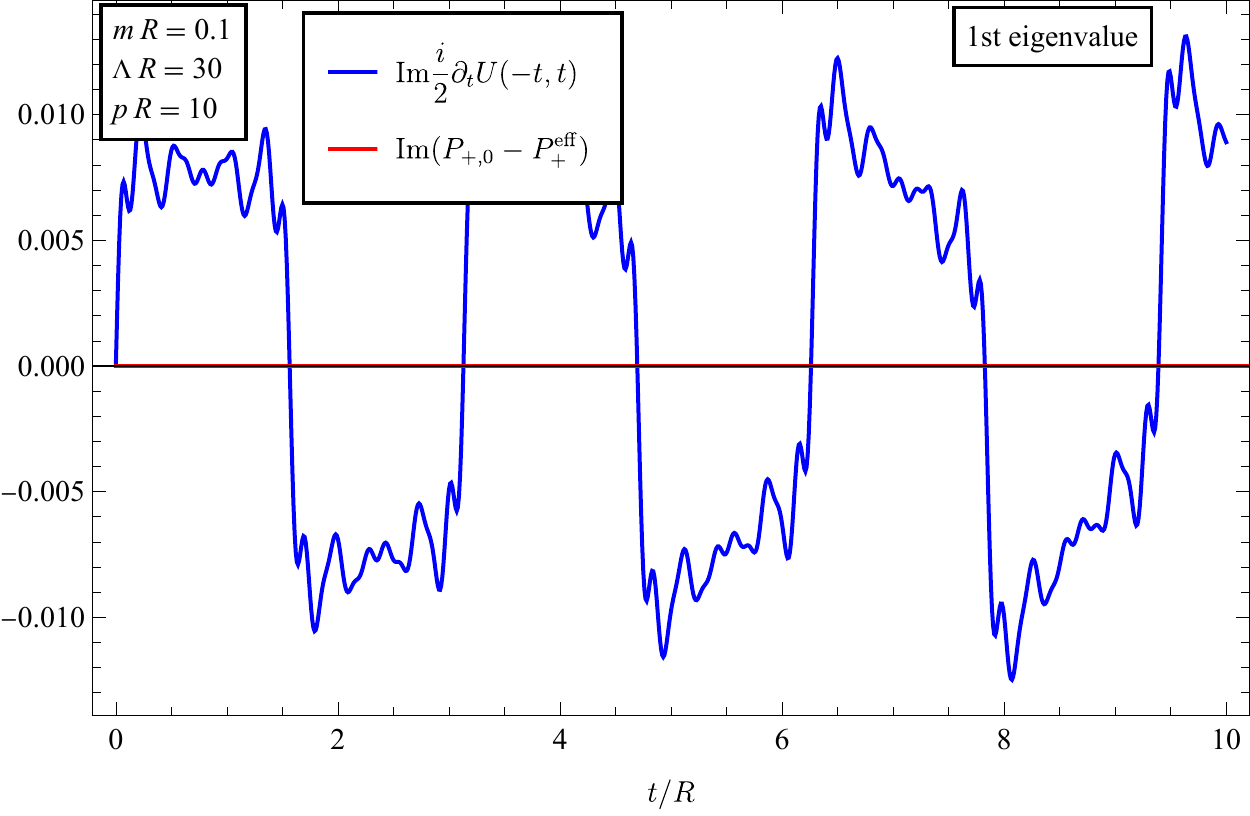}
\includegraphics[width=0.3\linewidth]{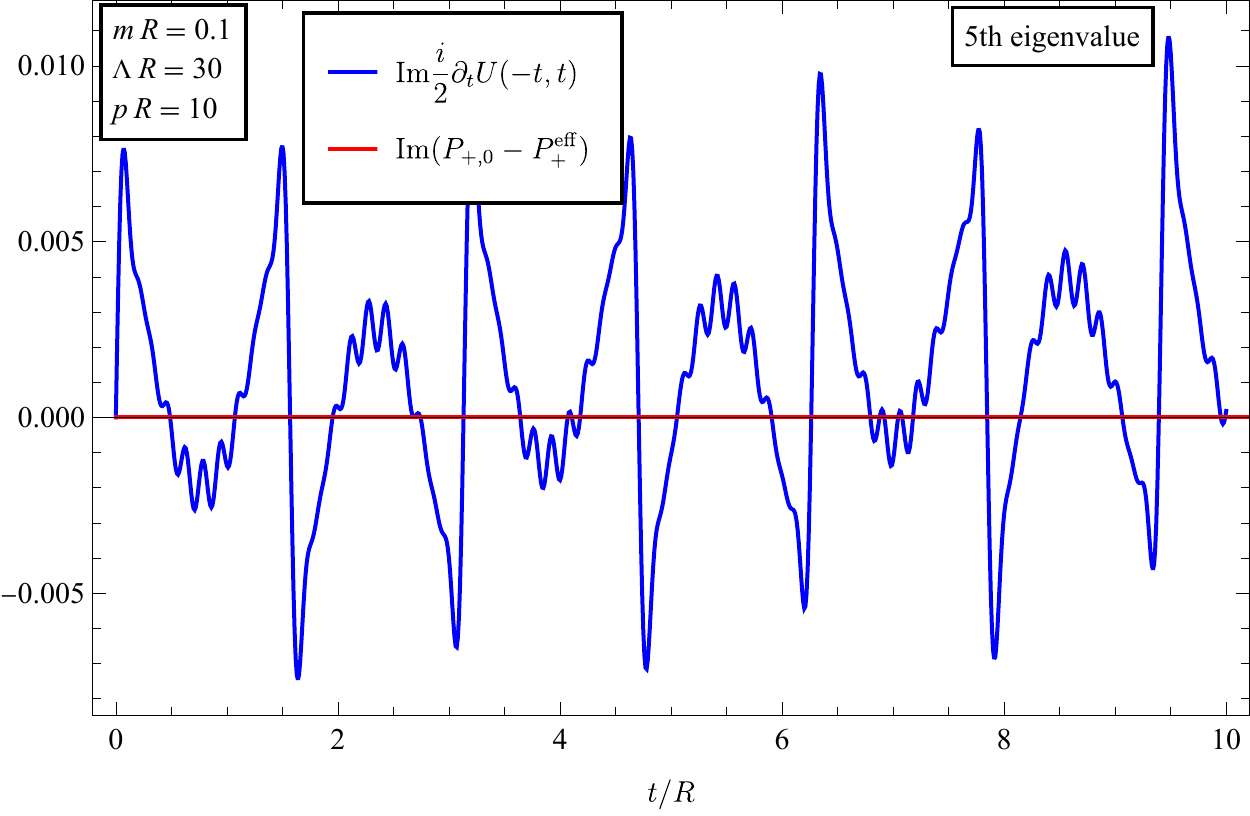}
\caption{Numeric depiction of $U(x^+,-x^+)$ and $\partial_{x^+} U(x^+,-x^+)$ in the range $E_h^{-1} \ll x^+ \ll E_l^{-1}$, in the case of the Ising model deformed by a mass term at large momentum.  Here, we see both the need for smearing over small range of $x^+$, as well as the fact that the average correctly computes the $Z$ factor and Hamiltonian, respectively. Strictly speaking one should divide $\partial_{x^+} U$ by $Z$ to get $P_+^{\rm eff}$ in the bottom four plots, but since $Z$ is so close to 1 in this case its effect is negligible.}
\label{fig:HfromAverages}
\end{figure}

\section{ Discussion \& Future directions}
\label{sec:Discussion}

The goal of this paper has been to develop the effective theory of the infinite momentum frame.  We claim that our formula (\ref{eq:HeffIntro}) gives a numeric method for computing the lightcone effective Hamiltonian in terms of the equal-time Hamiltonian, at least in $d=2$.  Conceptually, this formula is suggestive that 2d QFTs have a non-perturbative, volume independent, Lorentz invariant formulation.  In this formulation, QFT states are described in terms of a UV CFT chiral basis.  In this work we have checked (\ref{eq:HeffIntro}) for the full phase diagram of the Ising Field Theory.

However, ideally we would like to do better than a numeric method that relies on first computing (and potentially regularizing and renormalizing) an equal-time Hamiltonian.  Rather, we would like an explicit formulation of the effective action purely within the lightcone description.  In the low temperature phase of IFT, assuming the form (\ref{eq:BoostedVolumeDependenceHeffIntro}) such a description is possible because we can go to small volumes where $M^2_\sigma$ can be computed completely by using perturbation theory and working to leading order (where $\langle \sigma\rangle \sim g$).  In other words,
expanding our more general formula (\ref{eq:HeffIntro}) at second order in the coupling, we have that
\begin{equation}
\Ppeff = (P_+)_{ll} - (P_+)_{lh} \frac{1}{(P_+^{(0)})_{hh}} (P_+)_{hl} + \dots
\label{eq:Peff2ndOrderIntro}
\end{equation}
where $\dots$ indicates terms higher order than $\CO(g^2)$.  Compared to  (\ref{eq:HeffIntro}), this perturbative formula is vastly easier to evaluate, and moreover sidesteps the necessity of first obtaining and inverting the equal-time Hamiltonian; the second-order term can be rewritten in terms of a UV CFT two-point function of $\sigma$ evaluated between bra $\< l| $ and ket  $|l'\>$ states in the effective theory. Explicitly,
\begin{equation}
\<l | \frac{(P_+)_{lh}  (P_+)_{hl} }{(P_+^{(0)})_{hh}} | l'\> \propto \int_0^1 \frac{dr}{r} r^{h} \oint \frac{dx}{x}   \frac{dy}{y}  \frac{dz}{z} \frac{x^{p+h_l } }{z^{p-h_{l'}}} \< \CO_l(x) \sigma(1) \sigma(y,\bar{y}=\frac{r}{y}) \CO_{l'}(z)\>,
 \label{eq:HeffFromCorrIntro}
\end{equation}
where $\CO_l$ and $\CO_{l'}$ are operators that create LCT basis states in the lightcone effective theory.  In \cite{effH}, we will use this as a starting point for computing the effective Hamiltonian.

The punchline, as we show in \cite{effH}, is that (\ref{eq:HeffFromCorrIntro}) simplifies greatly in the infinite momentum limit. Indeed, we find a remarkably compact form for the effective interaction when $\eta \leq 0$:
\begin{equation}
\boxed{\Ppeff = m^2 \int dx^- \psi \frac{1}{2i\partial_-}\psi + g \< \sigma\> \int dx^- (\sigma(x) \sigma(\infty)-\mathbbm{1}).}
\label{eq:SimpleLCTHeff}
\end{equation}
Here $\sigma(\infty) \equiv \lim_{x \rightarrow \infty} (x^2)^{1/8} \sigma(x)$ is the relevant operator conformally mapped to the point at infinity, and $\sigma(x)$ is implicity at $x^+=0$.  This effective interaction manifestly has the correct transformation under boosts, even if one only transforms the holomorphic (i.e., $x^-$-dependent) half of $\sigma$,\footnote{Since our basis states are all chiral, the $x^+$ dependent half of $\sigma$ is essentially a spectator and does not contribute in the computation of the matrix elements.} since $\sigma_-(x^-) \rightarrow \lambda^{h_\sigma} \sigma_-(\lambda x^-)$, $\sigma_-(\infty) \rightarrow\lambda^{-h_\sigma} \sigma_-(\infty)$, and therefore $P_+ \rightarrow \lambda^{-1} P_+$.   In \cite{effH}, we will use this expression to solve the theory in the LCT basis.

Going forward, there are several interesting directions to explore.  One could try to compute other important observables in Ising Field Theory in LC; the S-matrix is a natural target  which has been the focus of recent work \cite{Correia:2022dyp,Gabai:2019ryw}.  An immediate question is what should be the correct effective Hamiltonian in the high temperature phase of IFT?
Our investigation using the Ramond sector for $\eta > 0$ might hint that chiral states with an odd number of fermions should be properly included in this phase, perhaps with the disorder operator playing a role.  More generally, one might apply (\ref{eq:HeffIntro}) to other 2d models with the hope of seeing structures similar to those present in (\ref{eq:SimpleLCTHeff}), or perhaps finding others that would flesh out the rules of the effective theory.  In particular, for the case of very relevant operators with $\Delta_i <1$, their respective vevs, $\langle \CO_i\rangle$, are well defined.  Does this indicate that terms similar to the second term in (\ref{eq:SimpleLCTHeff}) are generic for such operators?  Conversely, what happens for less relevant operators?  It would also be interesting to specifically examine supersymmetric (SUSY) 2d theories, where LC methods are often more compatible with SUSY\cite{Matsumura:1995kw,Hashimoto:1995jd,Lunin:1999ib,Fitzpatrick:2019cif}.  Moreover, a SUSY deformation will always have $\Delta_{\rm SUSY} \geq 1$, yet the vev $\langle \CO_{\rm SUSY} \rangle$ is either zero or finite, depending on whether SUSY is preserved or spontaneously broken, making SUSY an intriguing class of examples to study. Finally, what happens for $d>2$?  Does (\ref{eq:HeffIntro}) have a natural generalization for boosting QFTs in higher dimensions?

\section*{Acknowledgments}
We are grateful to Jo\~ao Penedones, Matthew Strassler and especially Matthew Walters   for  helpful discussions. ALF and EK are supported by the US Department of Energy Office of Science under Award Number DE-SC0015845, and the Simons Collaboration on the Non-Perturbative Bootstrap. YX is supported by a Yale Mossman Prize Fellowship in Physics.

\appendix
\section{Using the Ward Identity to Fix the Vacuum Energy}
\label{app:Vac}
In this paper, unless specified otherwise, we fix the vacuum energy as seen by the frame with momentum $p$ by imposing the Ward identity.  Specifically, we choose $|E_1\>$, to be the lightest eigenstate with momentum $p$, and demand that $E_{\rm vac}$ is the value that satisfies
\begin{equation}\label{wardId}
  \langle\Omega|T_{+-} |E_1\rangle = \frac{E_1-E_{\rm vac}-p}{E_1-E_{\rm vac}+p}\langle\Omega|T_{--} |E_1\rangle \, .
\end{equation}
The result in Fig \ref{fig:EvacSigOnly} suggests that the vacuum energy computed using the Ward Identity at large momentum should agree with that computed in the rest frame, so in principle we can determine the vacuum energy by running TCSA in the rest frame. But at large momentum there is a cancellation in $P_+ = \frac{1}{\sqrt{2}} \left( \sqrt{\mu^2+p^2} - p \right) \sim \CO(p^{-1})$ and a small truncation error in the TCSA vacuum energy measurement can cause a large deviation in the spectrum. In practice the Ward Identity gives us better accuracy in all the consistency checks.

The fact that (\ref{wardId}) refers to $T_{+-}$ means that the prescription always requires first solving the equal-time Hamiltonian $P_{+}$ and constructing $\Ppeff$ numerically using (\ref{eq:EffectiveHamiltonian}). Nevertheless the volume independence of $\Ppeff$ at large momentum allows us to go all the way down to $R\rightarrow 0$, where we can compute the vacuum energy analytically to the leading order in $m$ and $g$ in the rest frame.  In this case,
\begin{equation}\label{perturbativeEvac}
E_{\rm vac} = m^2 E_{\rm vac,\epsilon} + g^2 E_{\rm vac,\sigma}  \, .
\end{equation} 
Here, $E_{\rm vac,\epsilon}$ is the bubble diagram with a truncation on the two fermion intermediate state
\begin{equation}
E_{\rm vac,\epsilon} = -\sum_{q}^{q_{\rm cut}} \frac{|\<0|\epsilon|\psi_{q}\bar\psi_{q}\>|^2}{2q} = -\left( \frac{1}{2}H_{q_{\rm cut}-\frac{1}{2}}+\log 2 \right)
\end{equation}
where $q_{\rm cut} = \frac{1}{2}\left(\sqrt{\Lambda ^2+p^2}-p\right)$ and $H_k \equiv \sum_{j=1}^k \frac{1}{j}$.
$E_{\rm vac,\sigma}$ is finite, and we can lift the cutoff to infinity and obtain an integral on the two point correlation function
\begin{equation}
\label{EvacSigma}
E_{\rm vac,\sigma} = -\frac{1}{2} \oint \frac{dr}{r} r^h \oint \frac{dy}{y} \< \sigma(y,\bar{y})\sigma(1) \> = -\frac{2 \Gamma \left(\frac{1}{16}\right)^2 \Gamma \left(\frac{15}{8}\right)}{7 \Gamma
   \left(\frac{1}{8}\right) \Gamma \left(\frac{15}{16}\right)^2} = -8.00949
\end{equation}
where we substitute $\bar{y} = \frac{r}{y}$ and $h = \frac{1}{16}$.
With the analytic $E_{\rm vac}$, we can define $\Ppeff$ without solving the original $P_+$.

\section{Derivation of the correction to $P_+^\textrm{eff}$}\label{sec:effectiveHamiltonainCorrection}
\label{eq:PeffCorrections}
In this appendix, we provide a derivation of the correction to $\Ppeff$ mentioned in \eqref{eq:deltaPpeff}. The formula for $\delta \Ppeff$ is
\begin{equation}
	\delta \Ppeff =-\Ppeff Z^{-\frac{1}{2}}\Delta Y  Z^{-\frac{1}{2}}\Ppeff ,
  \label{eq:EffectiveHamiltonianCorrection}
\end{equation} 
where
\begin{equation}
\Delta Y \equiv P_{+}^{lh}\frac{1}{\left(P_{+}^{hh}\right)^{3}}P_{+}^{hl} \, .
\end{equation}

To derive this, start by including the $\CO(E_n^2)$ term in (\ref{eq:GeneralizedEvalEqn}):  
\begin{equation}
\left( Z^{\frac{1}{2}} \Ppeff Z^{\frac{1}{2}}\right) |\hat{n}\>_l = E_n Z |\hat{n}\>_l + E_n^2 \Delta Y |\hat{n}\>_l.
\end{equation}
As before, let $|\hat{n} \>_l \equiv Z^{-\frac{1}{2}} |\tilde{n}\>$.  Then,
\begin{equation}
\Ppeff |\tilde{n}\> = E_n | \tilde{n}\> + E_n^2 Z^{-\frac{1}{2}} \Delta Y Z^{-\frac{1}{2}} | \tilde{n}\> .
\label{eq:Eval2ndOrder}
\end{equation}
Since at leading order in $1/p$, $E_n$ is the eigenvalue of $\Ppeff$, we approximate $E_n^2$ in the correction term by the action of $\Ppeff$:
\begin{equation}
\Ppeff | \tilde{n} \> \approx E_n | \tilde{n}\> + \Ppeff Z^{-\frac{1}{2}} \Delta Y Z^{-\frac{1}{2}} \Ppeff  | \tilde{n}\>.
\end{equation}
Finally, we can write this as
\begin{equation}
(\Ppeff + \delta \Ppeff ) | \tilde{n} \> \approx E_n | \tilde{n}\>.
\end{equation}
Note that in the above derivation, the order of the operators when we replaced $E_n$s with $\Ppeff$s in (\ref{eq:Eval2ndOrder}) was not determined. We have found empirically that the above ordering works the best.

\bibliographystyle{JHEP}
\bibliography{refs}

\end{document}